\documentclass[a4paper,runningheads,orivec]{llncs}

\usepackage{mypreamble}

\title{\textsc{fbSAT}: Automatic Inference of Minimal Finite-State Models of Function Blocks Using SAT Solver
\thanks{Full version with appendices is available at \url{https://arxiv.org/abs/1907.03285}~\cite{fbSAT-full}}}
\titlerunning{\textsc{fbSAT}: Automatic Inference of Minimal FB Models Using SAT Solver}

\definecolor{orcidlogocolor}{HTML}{A6CE39}
\tikzset{
  orcidlogo/.pic={
    \fill[orcidlogocolor] svg{M256,128c0,70.7-57.3,128-128,128C57.3,256,0,198.7,0,128C0,57.3,57.3,0,128,0C198.7,0,256,57.3,256,128z};
    \fill[white] svg{M86.3,186.2H70.9V79.1h15.4v48.4V186.2z}
                 svg{M108.9,79.1h41.6c39.6,0,57,28.3,57,53.6c0,27.5-21.5,53.6-56.8,53.6h-41.8V79.1z M124.3,172.4h24.5c34.9,0,42.9-26.5,42.9-39.7c0-21.5-13.7-39.7-43.7-39.7h-23.7V172.4z}
                 svg{M88.7,56.8c0,5.5-4.5,10.1-10.1,10.1c-5.6,0-10.1-4.6-10.1-10.1c0-5.6,4.5-10.1,10.1-10.1C84.2,46.7,88.7,51.3,88.7,56.8z};
  }
}

\DeclareRobustCommand{\orcid}[1]{%
    \href{https://orcid.org/#1}{\mbox{\scalerel*{
        \begin{tikzpicture}[yscale=-1,transform shape]
            \pic{orcidlogo};
        \end{tikzpicture}
    }{|}}}
}

\author{%
    \textsuperscript{\orcid{0000-0002-4636-2379}}%
    Konstantin Chukharev\inst{1,2}
    \and
    \textsuperscript{\orcid{0000-0002-6417-6254}}%
    Daniil Chivilikhin\inst{1}
}
\authorrunning{K.~Chukharev \and D.~Chivilikhin}

\institute{Computer Technologies Laboratory, ITMO University, St.~Petersburg, Russia \and JetBrains Research, St.~Petersburg, Russia\\
\email{\{kchukharev,chivdan\}@itmo.ru}}

\let\oldbibliography\bibliography
\def\bibliography#1{}
\bibliography{bib}
\let\bibliography\oldbibliography
\undef\oldbibliography

\newcommand{\ifappendices}[2]{#1}

\begin{document}

\renewcommand{\onlyinsubfile}[1]{}
\renewcommand{\notinsubfile}[1]{#1}

\mainmatter

\maketitle

\begin{abstract}
Finite-state models are widely used in software engineering, especially in control systems development.
Commonly, in control applications such models are developed manually, hence, keeping them up-to-date requires extra effort.
To simplify the maintenance process, an automatic approach may be used, allowing to infer models from behavior examples and temporal properties.
As an example of a specific control systems development application we focus on inferring finite-state models of function blocks (FBs) defined by the IEC~61499 international standard for distributed automation systems.

In this paper we propose a method for FB model inference from behavior examples based on reduction to Boolean satisfiability problem (SAT).
Additionally, we take into account linear temporal properties using counterexample-guided synthesis.
We also present the developed tool \textsc{fbSAT} which implements the proposed method, and evaluate it in two case studies: inference of a finite-state model of a Pick-and-Place manipulator, and reconstruction of randomly generated automata.
In contrast to existing approaches, the suggested method is more efficient and produces finite-state models minimal \emph{both} in terms of number of states and guard conditions complexity.

\keywords{SAT \and Finite-state automata \and LTL \and Model checking \and Counterexample\-/guided inductive synthesis \and Function blocks \and IEC~61499}
\end{abstract}

\section{Introduction}%
\label{sec:introduction}

The non-trivial process of industrial control system development may be reduced to the development of a finite-state automaton or a system of interconnected automata.
The behavior of the controller may be represented using the deterministic finite-state model, allowing to describe how the system reacts to input events and which output actions it produces. Such models are extensively used in program testing \cite{model-testing,TESTOR} and verification \cite{buzhinsky-tii,RINGA}.
One practical example of finite-state model usage is the international standard for distributed automation systems development IEC~61499~\cite{vyatkin-tii}, which defines the control systems as networks of interconnected function blocks (FBs), specified by their \emph{interfaces} and implementations (\emph{control algorithms}).
Since the standard uses an event-driven execution model, the FB interface contains input/output events in addition to input/output data.

In practice, most finite-state models for control applications are developed \emph{manually} \--- this is a tedious and error-prone approach. Furthermore, there is the problem of maintaining these models to be up-to-date and consistent during the changes in system parameters, architecture, and logic.
An alternative to the manual process is \emph{automatic} synthesis from given execution scenarios and/or temporal properties~\cite{heule2010,bosy,efsm-tools,ulyantsev-lata,giantamidis-tripakis,petrenko,g4ltl-st,smetsers-lata}.
Inferred models can be used for model-based testing, verification and can even replace the original controller.

The contributions of this paper are the following.
\begin{enumerate}[leftmargin=1.5pc,topsep=2pt,itemsep=2pt]
    \item We propose and describe a method for automatic inference of minimal finite-state FB models, which allows \emph{simultaneously} and \emph{efficiently} accounting for (1)~behavior examples, (2)~LTL properties, and (3)~minimality of synthesized automata both in terms of number of states and guard conditions complexity.

    \item We present a tool \textsc{fbSAT} implementing the proposed method, and evaluate it in two case studies.
\end{enumerate}
Though our approach is implemented for FB model identification, it can be applied for inference of other types of state machines.

\section{Problem Statement}%
\label{sec:problem-statement}

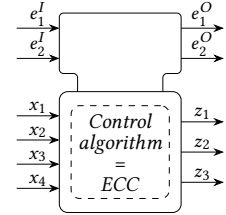
\begin{wrapfigure}{R}{0.23\textwidth}
    \vspace{-\intextsep}
    \centering
    \begin{adjustbox}{width=\linewidth}
        \begin{tikzpicture}[
    >={Stealth[]},
    auto,
]
    \def\mywidth{2}
    \def\myheighttop{1}
    \def\myheighttopgap{0.3}
    \def\myheightbot{\mywidth}
    \def\myarrowlength{0.7}

    \node[draw, dashed, rectangle, rounded corners, anchor=center, align=center] (center) at (\mywidth/2,-\mywidth/2)
        {\textit{Control} \\[-2pt] \textit{algorithm} \\[-2pt] = \\[-2pt] \textit{ECC}};
    \draw[rounded corners]
        (0,0) rectangle (\mywidth,-\myheightbot);
    \draw[-,rounded corners=2pt]
        (\myheighttopgap,0)
        -- ++(0,\myheighttopgap)
        -- ++(-\myheighttopgap,0)
        -- ++(0,\myheighttop)
        -- ++(\mywidth,0)
        -- ++(0,-\myheighttop)
        -- ++(-\myheighttopgap,0)
        -- ++(0,-\myheighttopgap)
    ;
    \path[inner sep=1pt, font=\small]
        (0,\myheighttopgap+0.75*\myheighttop) [<-]edge node[above] {$e^I_1$} +(-\myarrowlength,0)
        (0,\myheighttopgap+0.25*\myheighttop) [<-]edge node[above] {$e^I_2$} +(-\myarrowlength,0)
        (0,-0.2*\myheightbot) [<-]edge node[above] {$x_1$} +(-\myarrowlength,0)
        (0,-0.4*\myheightbot) [<-]edge node[above] {$x_2$} +(-\myarrowlength,0)
        (0,-0.6*\myheightbot) [<-]edge node[above] {$x_3$} +(-\myarrowlength,0)
        (0,-0.8*\myheightbot) [<-]edge node[above] {$x_4$} +(-\myarrowlength,0)
        (\mywidth,\myheighttopgap+0.75*\myheighttop) [->]edge node[above] {$e^O_1$} +(\myarrowlength,0)
        (\mywidth,\myheighttopgap+0.25*\myheighttop) [->]edge node[above] {$e^O_2$} +(\myarrowlength,0)
        (\mywidth,-0.25*\myheightbot) [->]edge node[above] {$z_1$} +(\myarrowlength,0)
        (\mywidth,-0.50*\myheightbot) [->]edge node[above] {$z_2$} +(\myarrowlength,0)
        (\mywidth,-0.75*\myheightbot) [->]edge node[above] {$z_3$} +(\myarrowlength,0)
    ;
\end{tikzpicture}%
    \end{adjustbox}
    \vskip-\abovecaptionskip
    \vspace{2pt}
    \caption{Function block example}%
    \label{fig:fb}
    \vskip-\belowcaptionskip
\end{wrapfigure}

A function block (FB) (\cref{fig:fb}) is characterized by its \emph{interface} and \emph{control algorithm}.
The interface defines input/output events (sets $\SetInputEvents$ and~$\SetOutputEvents$) and input/output variables (sets $\SetInputVariables$ and~$\SetOutputVariables$) which can be Boolean, integer or real-valued.
In this paper we consider Boolean input/output variables only.
The control algorithm is represented by a Moore finite-state machine, extended with guard conditions, and called \emph{execution control chart} (ECC).
Later we will refer to such a machine simply as \emph{an automaton}.
A complete formal definition of an ECC can be found in~\cite{dubinin-2006}.
We use a simplified one:
an automaton $\Automaton$ is a tuple $(\SetStates, \InitialState, \SetInputVariables, \SetOutputVariables, \SetInputEvents, \SetOutputEvents, \TransitionFunction*, \OutputEventFunction*, \OutputFunction*)$, where
$\SetStates$~is a set of states,
$\InitialState \in \SetStates$ \--- initial state,
$\SetInputVariables$~and $\SetOutputVariables$ \--- sets of input/output variables,
$\SetInputEvents$~and $\SetOutputEvents$ \--- sets of input/output events.
Each state has an associated output event and an \emph{algorithm} \--- function that changes values of output variables~$\SetOutputVariables$.
In this paper we consider algorithms of form \(\boolvec{\SetOutputVariables} \to \boolvec{\SetOutputVariables}\), where each output variable only depends on its previous value (assuming that initially it is \texttt{False}).
Each transition has an associated input event and a \emph{guard condition} \--- Boolean function that indicates the possibility to follow the transition. In this paper we consider guard conditions to be Boolean functions over input variables \(\SetInputVariables\), \ie of form \(\boolvec{\SetInputVariables} \to \Bool\).
An automaton $\Automaton$ is a finite-state transducer \--- it accepts \emph{input actions} and produces \emph{output actions}, while keeping track of output variable values.
Transition function \(\TransitionFunction* : \SetStates \times \SetInputEvents \times \boolvec{\SetInputVariables} \to \SetStates\) defines the state in which the automaton finishes processing an input action.
Output event function \(\OutputEventFunction* : \SetStates \times \SetInputEvents \times \boolvec{\SetInputVariables} \to (\SetOutputEvents \union \Set{\varepsilon})\) defines the output event emission rule.
Output function \(\OutputFunction* : \SetStates \times \SetInputEvents \times \boolvec{\SetInputVariables} \times \boolvec{\SetOutputVariables} \to \boolvec{\SetOutputVariables}\) defines the changes in output variable values.
Note that the automaton may not react to some input actions, \ie it may stay in the same state and not produce an output action. In that case \(\TransitionFunction*(q,e,\bar{x}) = q\), \(\OutputEventFunction*(q,e,\bar{x}) = \varepsilon\), and \(\OutputFunction*(q,e,\bar{x},\bar{z}) = \bar{z}\) ($q \in \SetStates, e \in \SetInputEvents, \bar{x} \in \boolvec{X}, \bar{z} \in \boolvec{Z}$).

An \emph{execution scenario} \(\SetScenarios\) is a sequence of \emph{elements} \(s_i = \Pair{ \InputAction, \OutputAction }\), where each element consists of an input action \(\InputAction\) and an output action \(\OutputAction\).
An input action is a pair of an input event $e^I \in \SetInputEvents$ and a tuple of input variable values $\bar{x} = \Vector{x_1, \dotsc, x_{\card{\SetInputVariables}}}$ $(x_i \in \SetInputVariables)$ later called \emph{input}, whereas an output action is a pair of an output event $e^O \in \SetOutputEvents \union \{\varepsilon\}$ and a tuple of output variable values $\bar{z} = \Vector{z_1, \dotsc, z_{\card{\SetOutputVariables}}}$ ($z_i \in \SetOutputVariables$) later called \emph{output}. An \emph{empty} output event $\varepsilon$ is necessary to represent the absence of an output action, \eg in case when an automaton does not react to an input action.
A \emph{positive scenario} is an execution scenario representing a desired behavior of an automaton.
Commonly, such scenarios are obtained by simulating an existing model (in a simulation tool, such as Matlab, nxtSTUDIO~\cite{nxtstudio}, \emph{etc}), or by interacting with a real control system.
An example of a set of three scenarios $\SetScenarios = \Set{s_1, s_2, s_3}$ is shown below:
\begin{equation}
    \label{eq:example-scenarios}
    \begin{aligned}
        & s_1 = [
            \Pair{\ActionTT{R}{00}, \ActionMT{\varepsilon}{0}};
            \Pair{\ActionTT{R}{01}, \ActionTT{B}{1}};
            \Pair{\ActionTT{R}{00}, \ActionMT{\varepsilon}{1}};
            \Pair{\ActionTT{R}{01}, \ActionTT{B}{0}}
        ], \\[-2pt]
        & s_2 = [
            \Pair{\ActionTT{R}{00}, \ActionMT{\varepsilon}{0}};
            \Pair{\ActionTT{R}{10}, \ActionTT{A}{0}};
            \Pair{\ActionTT{R}{00}, \ActionMT{\varepsilon}{0}};
            \Pair{\ActionTT{R}{01}, \ActionTT{B}{1}}
        ], \\[-2pt]
        & s_3 = [
            \Pair{\ActionTT{R}{00}, \ActionMT{\varepsilon}{0}};
            \Pair{\ActionTT{R}{10}, \ActionTT{A}{0}};
            \Pair{\ActionTT{R}{10}, \ActionTT{A}{0}}
        ].
    \end{aligned}
\end{equation}
An automaton is said to \emph{satisfy} a scenario if, while sequentially receiving input actions from scenario elements, the automaton produces exactly the same sequence of output actions as in the scenario.

An \emph{LTL specification}~$\LTLSpec$ is a set of \emph{LTL formulas} describing the temporal properties of a finite-state model.
An LTL formula is an expression which may contain propositional variables (in our case \--- input/output events/variables of the ECC), logical operators (\eg $\land, \lor, \neg, \rightarrow$), and temporal operators (\eg \temp{X} \--- \enquote{next}, \temp{U} \--- \enquote{until}, \temp{G} \--- \enquote{always}, \temp{F} \--- \enquote{eventually}).
An LTL specification can be verified using a model checker tool, which produces a counterexample for each violated LTL formula. We convert each counterexample into a \emph{negative scenario} represening an undesired behavior. We describe this in details in~\cref{sub:negative-scenarios}.

Ultimately, the problem addressed in this paper is to find the most \emph{general} automaton that satisfies all positive scenarios $\SetPositiveScenarios$ and complies with the given LTL specification~$\LTLSpec$.
Commonly, generalization of models is achieved through minimizing their number of states and/or transitions \cite{heule2010,efsm-tools,petrenko}.
In this work we additionally explicitly consider complexity of guard conditions: generalization is achieved by minimizing the sought automaton \emph{both} in terms of the number of states and the complexity of its guard conditions, measures as the total number of vertices in parse trees of corresponding Boolean formulas.

\vspace{-4pt}
\section{Related Work}%
\label{sec:related-work}
\vspace{-2pt}

There exists a large body of work on SAT-based synthesis of circuits, bit-vector programs, domain-specific programs, etc.
However, in this work we are interested specifically in synthesis of finite-state machines: state-based models are comprehensible, their formal verification is relatively simple, and they can be directly used in control applications for controller logic implementation, e.g. in Matlab/Stateflow, nxtSTUDIO~\cite{nxtstudio}. 

The problem of finding a minimal deterministic finite-state machine from behavior examples is known to be NP-complete~\cite{gold}, and the complexity of the LTL synthesis problem is double exponential in the length of the LTL specification~\cite{rosner-phd}.
Despite this, synthesis of various types of finite-state models from behavior examples and/or formal specification has been addressed by many researchers including~\cite{heule2010,efsm-tools,zakirzyanov2019,buzhinsky-tii,bosy,tsarev-egorov-gecco,giantamidis-tripakis,petrenko,petrenko2,neider,g4ltl-st,regular-inference,smetsers-lata} with methods based on heuristic state merging, evolutionary algorithms and SAT/SMT-solvers.
In the context of this paper we are interested in exact methods, so we direct our attention to SAT-based methods.

Extended Finite-State Machine (\emph{EFSM}) is the model most similar to the ECC considered in this paper: it combines a Mealy and a Moore automaton extended with conditional transitions. Transitions are labeled with input events and Boolean formulas over the input variables, and automaton states have associated sequences of output actions.
Several approaches based on translation to SAT~\cite{efsm-tools,walkinshaw} have been proposed for inferring EFSMs from behavior examples and LTL properties.
In~\cite{efsm-tools} LTL properties are accounted for via an iterative counterexample prohibition approach.

The BoSy tool~\cite{bosy,not-bosy} implements bounded synthesis of a \emph{transition system} (a type of automaton similar to EFSM and ECC) from given LTL properties. Synthesis is \emph{bounded} in the sense that the number of states does not exceed a given bound. Apart from the SAT-based approach, a more efficient solution based on a Quantified SAT~(QSAT) encoding is developed.
Transition systems inferred with the SAT encoding are \emph{explicit} (guard conditions include all input variables), whereas the QSAT encoding yields \emph{symbolic} models (guard conditions are Boolean formulas over input variables).
BoSy ensures that found solutions are minimal \wrt the number of states, however it does not allow minimizing guard conditions, which tend to be large and incomprehensible.
An approach to make generated solutions simpler is suggested in~\cite{bounded-cycle}, where the SAT encoding is augmented with constraints for minimizing the number of cycles in the transition system. However, guard conditions complexity is not addressed. Furthermore, BoSy does not support behavior examples. Though they can be modeled with LTL formulas, this approach is inefficient even for behavior examples of moderate size.
Other LTL synthesis techniques, \eg G4LTL-ST~\cite{g4ltl-st} and Strix~\cite{strix}, have the same drawbacks in application to the considered problem: no guard conditions minimization and lack of support for behavior examples.

In~\cite{fbCSP}, the \textsc{fbCSP} method is proposed for inferring an FB model from execution scenarios via a translation to the Constraint Satisfaction Problem~(CSP).
However, \textsc{fbCSP} has the following restrictions.
Guard conditions are generated in \emph{complete} form \--- corresponding Boolean formulas depend on \emph{all} input variables. Such models do not generalize to unseen data.
This is countered by greedy guard conditions minimization, but it does not guarantee the result minimality.
In~\cite{chivilikhin-18} \textsc{fbCSP} is extended with a counterexample prohibition procedure similar to~\cite{efsm-tools} to account for LTL properties. Guard conditions are represented with fixed-size conjunctions of positive/negative literals of input variables. The drawback of this approach is that it is inefficient models when temporal properties are insufficiently covered with behavior examples.

In~\cite{chivilikhin-19} the approach of \textsc{fbCSP} is developed further:
on the first stage, a base model is inferred with a translation to SAT, and on the second stage its guard conditions are minimized via a CSP encoding, in which guard condition Boolean formulas are represented with parse trees. By introducing a total bound on the number of nodes in these parse trees and solving a series of CSP problems, the method finds a model with minimal guard conditions \wrt the base model identified on the first stage.
Global minimality of guard conditions is not guaranteed due to the two-stage implementation: minimal guards may correspond to another base model, not the one found on the first stage.
The same argument applies against any approach based on state machine minimization~\cite{klenze}.
In addition, LTL properties are not supported.

Overall, none of the existing methods allow \emph{simultaneously} and \emph{efficiently} accounting for (1)~behavior examples, (2)~LTL properties, and (3)~minimality of synthesized automata in terms of both number of states and guard conditions complexity. The approach proposed in this paper extends~\cite{chivilikhin-19} and contributes to the state-of-the-art in SAT-based state machine synthesis: it supports positive behavior examples, realizes counterexample-guided synthesis to account for LTL properties, and produces models minimal both in terms of the number of states and guard conditions complexity.

\subfilebiblio

\vspace{-4pt}
\section{Proposed Approach}%
\label{sec:proposed-approach}
\vspace{-2pt}

In this section we develop our approach for inferring minimal FB models from a given set of positive scenarios and an LTL specification.
In \cref{sub:scenario-tree} we describe a convenient storage structure for execution scenarios \--- \emph{scenario tree}.
In \cref{sub:negative-scenarios} we describe the process of verifying an LTL specification using a model checker tool, which produces a counterexample for each violated LTL formula. Obtained counterexamples are converted into \emph{negative scenarios} representing the undesired behavior, which must be prohibited.
In \cref{sub:model-inference} we describe the reduction of the FB model inference problem to SAT\@.
In \cref{sub:minimal-model-inference} we describe the process of inferring a minimal FB model both in terms of the number of states and guard conditions complexity.

\subsection{Scenario Tree Construction}%
\label{sub:scenario-tree}

A \emph{scenario tree}~$\Tree$ is a prefix tree built from the given scenarios~$\SetScenarios$.
Before the scenario tree construction, we prepend each scenario with an auxiliary element consisting only of an output action $\Action{\varepsilon}{\Vector{0 \dots 0}}$. By this we ensure that all scenarios have a common prefix.
Each tree node and its incoming edge correspond to a scenario element: a node is marked with an output action, and an edge is marked with an input action.

\begin{figure}[!htb]
    \centering
    \begin{adjustbox}{max height=\dimexpr\pagegoal-\pagetotal-\baselineskip-\abovecaptionskip-\belowcaptionskip\relax, min width=0.4\textwidth}
        \begin{tikzpicture}[
    auto,
    on grid,
    every state/.style={inner sep=0pt},
    every label/.style={inner sep=1pt},
]
    \def\VGapOne{0.9cm}
    \def\VGapTwo{0.6cm}
    \def\HGap{2cm}
    \def\HGapOne{\HGap}
    \def\HGapTwo{\HGap}
    \def\BendAngle{20} %

    \node[state,label={above:$v_{1}$}] (v1) at (0,0)                                  {\ActionMT{\varepsilon}{0}};
    \node[state,label={above:$v_{2}$}] (v2) [right=\HGap of v1]                       {\ActionMT{\varepsilon}{0}};
    \node[state,label={above:$v_{3}$}] (v3) [above right=\VGapOne and \HGapOne of v2] {\ActionTT{B}{1}};
    \node[state,label={above:$v_{4}$}] (v4) [right=\HGap of v3]                       {\ActionMT{\varepsilon}{1}};
    \node[state,label={above:$v_{5}$}] (v5) [right=\HGap of v4]                       {\ActionTT{B}{0}};
    \node[state,label={above:$v_{6}$}] (v6) [below right=\VGapOne and \HGapOne of v2] {\ActionTT{A}{0}};
    \node[state,label={above:$v_{7}$}] (v7) [above right=\VGapTwo and \HGapTwo of v6] {\ActionMT{\varepsilon}{0}};
    \node[state,label={above:$v_{8}$}] (v8) [right=\HGap of v7]                       {\ActionTT{B}{1}};
    \node[state,label={above:$v_{9}$}] (v9) [below right=\VGapTwo and \HGapTwo of v6] {\ActionTT{A}{0}};

    \path[-{Stealth[]}, shorten >=1pt]
        (v1) edge                        node                      {\ActionTT{R}{00}} (v2)
        (v2) edge[bend left=\BendAngle]  node[sloped,anchor=south] {\ActionTT{R}{01}} (v3)
        (v3) edge                        node                      {\ActionTT{R}{00}} (v4)
        (v4) edge                        node                      {\ActionTT{R}{01}} (v5)
        (v2) edge[bend right=\BendAngle] node[sloped,anchor=north] {\ActionTT{R}{10}} (v6)
        (v6) edge[bend left=\BendAngle]  node[sloped,anchor=south] {\ActionTT{R}{00}} (v7)
        (v7) edge                        node                      {\ActionTT{R}{01}} (v8)
        (v6) edge[bend right=\BendAngle] node[sloped,anchor=north] {\ActionTT{R}{10}} (v9)
    ;

    \begin{scope}[every text node part/.style={align=left}]
    \node at (v1.west) [left, inner sep=0pt] {%
        $\card{\SetTreeNodes} = 9,~ \rho = v_1,$ \\
        $\SetInputEvents = \Set{\texttt{R}},~ \SetOutputEvents = \Set{\texttt{A}, \texttt{B}},$ \\
        $\card{\SetInputVariables} = 2,~ \card{\SetOutputVariables} = 1,$ \\
        $\SetTreeInputs = \Set{
            \Vector{\texttt{00}},
            \Vector{\texttt{01}},
            \Vector{\texttt{10}},
            \Vector{\texttt{11}}
        }$
    };
    \end{scope}
\end{tikzpicture}%
%
    \end{adjustbox}
    \vskip-\abovecaptionskip
    \caption{Scenario tree constructed from scenarios~(\ref{eq:example-scenarios})}%
    \label{fig:scenario-tree}
    \vskip-\belowcaptionskip
\end{figure}
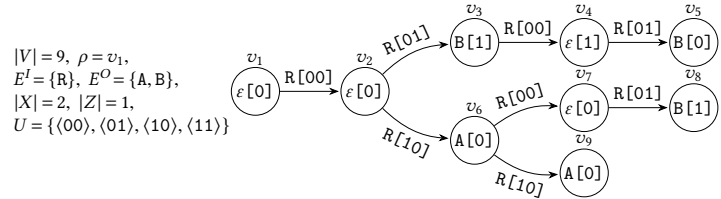

Further in this paper, we will refer to the key features of a scenario tree as follows:
$\SetTreeNodes$~is a set of tree nodes;
$\TreeRoot \in \SetTreeNodes$ \--- root of the tree;
$\tp{v} \in \SetTreeNodes$ \--- parent of node $v \neq \TreeRoot$;
$\tie{v} \in \SetInputEvents$ \--- input event on the incoming edge of node~$v \neq \TreeRoot$;
$\toe{v} \in \SetOutputEvents \union \Set{\varepsilon}$ \--- output event in node~$v$, where $\varepsilon$~is an~empty event;
$\SetTreeNodesActive = \Set{v \in \SetTreeNodes \setminus \Set{\TreeRoot} \given \toe{v} \neq \varepsilon}$ \--- set of \emph{active} tree nodes;
$\SetTreeNodesPassive = \Set{v \in \SetTreeNodes \setminus \Set{\TreeRoot} \given \toe{v} = \varepsilon}$ \--- set of \emph{passive} tree nodes;
$\SetTreeInputs \subseteq \boolvec{\SetInputVariables}$ \--- set of \emph{inputs} encountered in scenarios;
$\tin{v} \in \SetTreeInputs$ \--- input on the incoming edge of node~$v \neq \TreeRoot$;
$\tov{v,z} \in \Bool$ \--- value of output variable~$z$ in node~$v$.
The root~$\TreeRoot$ has no parent, thus~$\tp{\TreeRoot}$, $\tie{\TreeRoot}$, and~$\tin{\TreeRoot}$ are undefined.
A \emph{positive scenario tree}~$\PositiveTree$ is a scenario tree built from positive scenarios~$\SetPositiveScenarios$.
An example of a scenario tree constructed from scenarios~(\ref{eq:example-scenarios}) is shown in \cref{fig:scenario-tree}.

\subsection{LTL Verification, Counterexamples, Negative Scenarios}%
\label{sub:negative-scenarios}

An LTL specification can be verified using a model checker tool, which produces a \emph{counterexample} for each violated LTL formula.
We use a symbolic model checker NuSMV~\cite{NuSMV}.
For safety properties, a counterexample is a finite sequence of \emph{execution states}.
For liveness properties, a counterexample is an infinite but periodic sequence of states, which can be represented as a finite prefix followed by a cycle~\cite{clarke1999model}.

\begin{figure}[!htb]
    \centering
    \begin{adjustbox}{max width=\textwidth}
        \begin{tikzpicture}[
    >={Stealth[]},
    shorten >=1pt,
    auto,
    on grid,
    initial text={},
    label distance=0pt,
    every state/.style={inner sep=0pt, font=\large},
    every label/.style={inner sep=1pt, anchor=center},
]
    \def\HGap{1.8cm}
    \def\VGap{1.1cm}
    \def\HHGap{2.1cm}

    \node[initial,state,label={-100:$\StateOutputEvent{q_1} = \varepsilon$}] (q1) at (0,0)                  {$q_1$};
    \node[state,label={ 30:$\StateOutputEvent{q_2} = \texttt{A}$}] (q2) [above right=\VGap and \HGap of q1] {$q_2$};
    \node[state,label={-30:$\StateOutputEvent{q_3} = \texttt{A}$}] (q3) [below right=\VGap and \HGap of q1] {$q_3$};
    \node[state,label={-80:$\StateOutputEvent{q_4} = \texttt{A}$}] (q4) [below right=\VGap and \HGap of q2] {$q_4$};
    \node[state,label={above:$\StateOutputEvent{q_5} = \texttt{B}$}] (q5) [right=\HHGap of q4]              {$q_5$};

    \path[->]
        (q1) edge               node[sloped,anchor=south]      {$\texttt{R} \,\&\, x_1$}      (q2)
        (q1) edge               node[sloped,anchor=north,swap] {$\texttt{R} \,\&\, \neg x_1$} (q3)
        (q2) edge               node[sloped,anchor=south]      {$\texttt{R} \,\&\, x_1$}      (q4)
        (q3) edge[bend left=15] node[sloped,anchor=south]      {$\texttt{R} \,\&\, x_1$}      (q2)
        (q2) edge[bend left=15] node[sloped,anchor=south]      {$\texttt{R} \,\&\, \neg x_1$} (q3)
        (q4) edge               node[sloped,anchor=north,swap] {$\texttt{R} \,\&\, \neg x_1$} (q3)
        (q4) edge               node                           {$\texttt{R} \,\&\, x_1$}      (q5)
    ;

    \begin{scope}[every text node part/.style={align=left}]
    \node [left=3cm of q1] {%
        $\card{\SetStates} = 5,~ \InitialState = q_1$ \\
        $\SetInputEvents = \Set{\texttt{R}},~ \SetOutputEvents = \Set{\texttt{A}, \texttt{B}}$ \\
        $\card{\SetInputVariables} = 2,~ \card{\SetOutputVariables} = 1$ \\
    };
    \end{scope}
\end{tikzpicture}%
%
    \end{adjustbox}
    \vskip-\abovecaptionskip
    \caption{Example automaton with a looping behavior}%
    \label{fig:looping}
    \vskip-\belowcaptionskip
\end{figure}
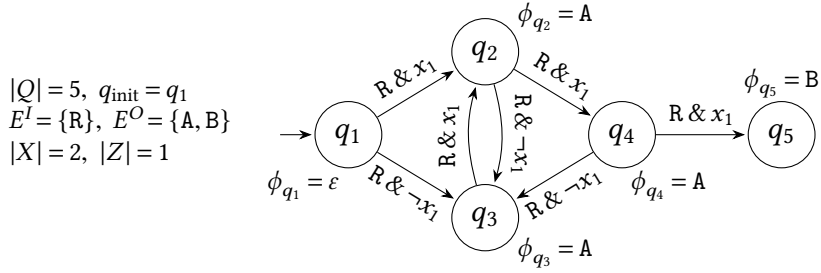

A \emph{negative scenario} is an execution scenario representing an undesired behavior.
We convert each counterexample into a negative scenario as follows.
Consider an automaton in~\cref{fig:looping} and two LTL properties: \(\LTLSpec = \Set{\temp{G}(e^{O\!\!} \neq \texttt{B}),\allowbreak \temp{F}(e^{O\!\!} = \texttt{B})}\).
Counterexample for the safety property $\LTLSpec_1$ is a finite sequence \(
    \ExecutionState{q_1}{\varepsilon}
    \raisebox{-2pt}{$\xrightarrow{\smash{\ActionTT{R}{1}}}$}
    \ExecutionState{q_2}{A}
    \raisebox{-2pt}{$\xrightarrow{\smash{\ActionTT{R}{1}}}$}
    \ExecutionState{q_4}{A}
    \raisebox{-2pt}{$\xrightarrow{\smash{\ActionTT{R}{1}}}$}
    \ExecutionState{q_5}{A}
\). Corresponding \emph{loopless} negative scenario: \(
    \Negative{s}_1 = [
        \Pair{\ActionTT{R}{1}, \texttt{A}};
        \Pair{\ActionTT{R}{1}, \texttt{A}};
        \Pair{\ActionTT{R}{1}, \texttt{A}}
    ]
\).
Counterexample for the liveness property $\LTLSpec_2$ is a finite prefix followed by a repeating cycle:
\[
    \overbraceeq[\textit{ }]{\textit{finite}}{\textit{prefix}}{
        \ExecutionState{q_1}{\varepsilon}
        \raisebox{-2pt}{$\xrightarrow{\ActionTT{R}{1}}$}
    }
    \overbrace{
        \tikz[remember picture, inner sep=0pt, outer sep=0pt, baseline, anchor=base]{
            \node (loop-start) {$\ExecutionState{q_2}{A}$};
        }
        \raisebox{-2pt}{$\xrightarrow{\ActionTT{R}{1}}$}
        \ExecutionState{q_4}{A}
        \raisebox{-2pt}{$\xrightarrow{\ActionTT{R}{0}}$}
        \ExecutionState{q_3}{A}
        \tikz[remember picture, inner sep=0pt, outer sep=0pt, baseline=-0.12ex, >={Stealth[]}, shorten >=1pt]{
            \draw[overlay, ->, densely dashed, rounded corners] (0,0)
                .. controls ++(0.7,0) and ++(0.7,0) .. (0,-0.35)
                to[out=180, in=-25, in looseness=0.4] (loop-start.south);
        }
        \raisebox{-2pt}{$\xrightarrow{\ActionTT{R}{1}}$}
        \raisebox{0pt}[0pt][\dp\strutbox+3mm]{}
    }^{\textit{cycle}}
    \ExecutionState{q_2}{A} .
\]
Corresponding \emph{looping} negative scenario: \(
    \Negative{s}_2 = [
        \underline{\Pair{\ActionTT{R}{1}, \texttt{A}}};\allowbreak
        \Pair{\ActionTT{R}{1}, \texttt{A}};\allowbreak
        \Pair{\ActionTT{R}{0}, \texttt{A}};\allowbreak
        \Pair{\ActionTT{R}{1}, \texttt{A}}
    ]
\), where the first element (underlined) is a beginning of a loop.

A \emph{negative scenario tree}~$\NegativeTree$ is a scenario tree built from negative scenarios~$\SetNegativeScenarios$.
We denote the set of all nodes that correspond to the last elements of loopless scenarios as~$\SetNegativeTreeNodesEnds$.
We augment the tree with special \emph{back edges} from the end to the beginning of each loop from looping scenarios. We denote the set of all nodes, which are connected with node $\negv$ via a back edge as~$\negtbe{\negv} \subseteq \SetNegativeTreeNodes$.
For example, for the negative scenario tree built from the scenario $\Negative{s}_2$ only: $\negtbe{\negv_5} = \Set{\negv_2}$ (indices are shifted by 1, since we prepend the scenario tree with an auxiliary root node $\NegativeTreeRoot = \negv_1$).
All other tree features are the same as defined in \cref{sub:scenario-tree}, but marked with a hat symbol, \eg $\negv \in \SetNegativeTreeNodes$, $\negtp{\negv}$, $\negtie{\negv}$.

\subsection{FB Model Inference Using SAT Solver}%
\label{sub:model-inference}

We propose a method for inferring an FB model based on the reduction to SAT\@.
The reduction consists in formally describing an automaton~$\Automaton$ of size~$C$ by constructing a Boolean formula that is satisfiable if and only if there exists an automaton which satisfies given positive scenarios~$\SetPositiveScenarios$ and does not satisfy given negative scenarios~$\SetNegativeScenarios$.
In order to encode non-Boolean variables with bounded domains we use a standard pairwise encoding (also known as \enquote{sparse} or \enquote{direct}~\cite{direct-encoding}) and Onehot+Binary~\cite{onehot-binary} encoding.
For all constraints presented not in CNF we apply a Tseytin transform.

The proposed reduction consists of four parts.
First, we encode the \emph{automaton structure} by declaring corresponding variables and constraints.
Second, we encode the \emph{positive} scenario tree~$\PositiveTree$ \emph{mapping} and enforce its satisfaction.
Third, we encode the \emph{guard conditions structure}, \ie the structure of parse trees of corresponding Boolean formulas, and declare cardinality constraints allowing to bound the guard conditions complexity.
Lastly, we encode the \emph{negative} scenario tree~$\NegativeTree$ \emph{mapping} and prohibit its satisfaction.

The goal is to infer an automaton with $\card{\SetStates} = C$ states.
We assume that each state has at most~$K$ outgoing transitions.
\(K = C \cdot \card{\SetInputEvents}\) is the safest minimum value that does not prohibit the inference of an automaton, which may happen for smaller values of $K$ due to over-constraining.
However, lowering this value greatly reduces the size of the reduction, which is likely to significantly increase the solving efficiency.
Further in this section we assume that $b \in \Bool = \Set{\top, \bot}$, $q \in \SetStates$, $k \in [1 \dd K]$, $e \in \SetInputEvents$, $u \in \SetTreeInputs$, $v \in \SetTreeNodes$, unless stated otherwise.

\subsubsection{Automaton Structure Encoding.}

Each state has an associated output event and an algorithm.
Variable \(\StateOutputEvent{q} \in \SetOutputEvents \union \Set{\varepsilon}\) denotes the output event in state $q$.
Variable \(\StateAlgorithm{q,z,b} \in \Bool\) represents the algorithm for the output variable $z$.

Each transition has an associated input event and a guard condition \--- Boolean function over input variables $\SetInputVariables$.
Variable \(\TransitionDestination{q,k} \in \SetStatesAux = \SetStates \union \Set{q_0}\) denotes the destination state of the $k$-th transition from state $q$.
\enquote{Transitions} to the auxiliary state $q_0 \notin \SetStates$ are called \emph{null}-transitions and represent the absence of a transition.
\WLoG[], we ensure that null-transitions have the largest indices: \((\TransitionDestination{q,k} = q_0) \implies (\TransitionDestination{q,k+1} = q_0)\).
Variable \(\TransitionInputEvent{q,k} \in \SetInputEvents \union \Set{\varepsilon}\) denotes the input event.
Only null-transitions are marked with an $\varepsilon$ input event: \((\TransitionDestination{q,k} = q_0) \iff (\TransitionInputEvent{q,k} = \varepsilon)\).
Variable \(\TransitionFiring{q,k,u} \in \Bool\) denotes whether a guard fires on input~$u$.
According to the IEC~61499 standard, each state has a transition priority: the automaton follows the \emph{first fired} transition or stays in the same state if no transition fired.
Variable \(\TransitionFirstFired{q,u} \in [0 \dd K]\) denotes the index of a transition which \emph{fires first} on input~$u$. \(\TransitionFirstFired{q,u} = 0\) denotes that no transition fires at all.
A transition fires first iff all previous transitions do not fire:
\(
    (\TransitionFirstFired{q,u} = k)
    \iff
    \TransitionFiring{q,k,u}
    \land
    \biglandnolim{1 \leq k' < k}
    (\neg\TransitionFiring{q,k',u})
\).

When the automaton in state $q$ processes an input action \(\Action{e}{u}\), it either (1)~goes to another state $q'$ or (2)~\emph{ignores} it (or rather, \emph{reacts by ignoring}) by staying in the state $q$.
Such behavior is represented by variable \(\ActualTransitionFunction{q,e,u} \in \SetStatesAux\), where \(\ActualTransitionFunction{q,e,u} = q_0\) denotes the second (2) case.
Note that in the first (1) case the automaton may go (through a \emph{loop}-transition) into the same state $q' = q$.

Additionally, we declare auxiliary symmetry-breaking constraints~\cite{ulyantsev-lata}, which force the automaton states to be enumerated in the order they are visited by the breadth-first search (BFS) algorithm launched from the initial state.
Variable $t^\text{bfs}_{q_i,q_j} \in \Bool$ (${q_i,q_j \in \SetStates}$) indicates the existence of a transition from $q_i$ to $q_j$: \(
    t^\text{bfs}_{q_i,q_j}
    \iff
    \biglornolim{k \in [1 \dd K]}
    (\TransitionDestination{q_i,k} = q_j)
\).
Variable $p^\text{bfs}_{q_j} \in \Set{q_1,\dotsc,q_{j-1}}$ ($j \in [2 \dd C]$) denotes the parent of the state $q_j$ in the BFS traverse tree: \(
    (p^\text{bfs}_{q_j} = q_i)
    \iff
    t^\text{bfs}_{q_i,q_j}
    \land
    \biglandnolim{r < i}
    (\neg t^\text{bfs}_{q_r,q_j})
\).
Actual BFS constraint is defined as follows:
\(
    (p^\text{bfs}_{q_j} = q_i)
    \implies
    \biglandnolim{r < i}
    (p^\text{bfs}_{q_{j+1}} \neq q_{r})
\).

\vspace{2cm}

\begin{wrapfigure}{R}{0.41\textwidth}
    \vspace{-\intextsep}
    \centering
    \begin{adjustbox}{width=\linewidth}
        \begin{tikzpicture}[
    >={Stealth[]},
    shorten >=1pt,
    auto,
    on grid,
    every state/.style={inner sep=0pt, font=\Large},
]
    \def\HGapAutomaton{0.4cm}
    \def\VGapAutomaton{1.7cm}
    \def\HGapTree{0.4cm}
    \def\VGapTree{1.5cm}
    \def\HGapBetween{1.4cm}
    \def\VGapBetween{1.1cm}
    \def\LineGapSmall{-5pt}
    \def\LineGapMore{-6pt}
    \definecolor{mylightgray}{RGB}{225,225,225}
    \definecolor{mydarkgray}{RGB}{180,180,180}

    \node[state] (qi)
        [fill=mylightgray]
        {$q_i$};
    \node[state] (qj)
        [fill=mydarkgray]
        [below left=\VGapAutomaton and \HGapAutomaton of qi]
        {$q_j$};
    \node[state] (v1)
        [fill=mylightgray]
        [above left=\VGapBetween and \HGapBetween of qi]
        {$v_1$};
    \node[state] (v2)
        [fill=mylightgray]
        [below left=\VGapTree and \HGapTree of v1]
        {$v_2$};
    \node[state] (v3)
        [fill=mydarkgray]
        [below left=\VGapTree and \HGapTree of v2]
        {$v_3$};

    \node at (v1.west) [left]
        {\(\toe{v_1} = o_1\)};
    \node at (v2.west) [left]
        {\(\begin{aligned}
            \tie{v_2} &= e_1 \\[\LineGapSmall]
            \tin{v_2} &= u_1 \\[\LineGapSmall]
            \toe{v_2} &= o_2 = \varepsilon %
        \end{aligned}\)};
    \node at (v3.west) [left]
        {\(\begin{aligned}
            \tie{v_3} &= e_2 \\[\LineGapSmall]
            \tin{v_3} &= u_2 \\[\LineGapSmall]
            \toe{v_3} &= o_3 %
        \end{aligned}\)};

    \node[draw,rounded corners,inner sep=3pt]
        (tree-label)
        at ([xshift=-16mm,yshift=10mm] v2)
        {Tree};
    \node[draw,rounded corners,inner sep=3pt]
        (automaton-label)
        at ([xshift=5mm, yshift=-8mm] qj)
        {Automaton};

    \node at (qi.east) [right] {$\StateOutputEvent{q_i\!} = o_1$};
    \node at (qj.east) [right] {$\StateOutputEvent{q_j\!} = o_3$};

    \path[->]
        (qi) edge node[right] {
            \(\begin{aligned}
                & \ActualTransitionFunction{q_i,e_1,u_1\!} = q_0 \\[\LineGapMore]
                & \ActualTransitionFunction{q_i,e_2,u_2\!} = q_j
            \end{aligned}\)} (qj)
        (v1) edge node[sloped,above,font=\tiny] {$\Action{e_1}{u_1}$} (v2)
        (v2) edge node[sloped,above,font=\tiny] {$\Action{e_2}{u_2}$} (v3)
        (v1) edge[densely dashed,out=0,in=135] node[above right,font=\tiny] {$\Mapping{v_1\!} = q_i$} (qi)
        (v2) edge[densely dashed,out=0,in=180] node[pos=0,below right,font=\tiny] {$\Mapping{v_2\!} = q_i$} (qi)
        (v3) edge[densely dashed,out=0,in=180] node[pos=0,below right,font=\tiny] {$\Mapping{v_3\!} = q_j$} (qj)
    ;

\end{tikzpicture}%
%
    \end{adjustbox}
    \vskip-\abovecaptionskip
    \caption{Tree\--to\--automaton mapping \mbox{example}}%
    \label{fig:tree-automaton-mapping}
    \vskip-\belowcaptionskip
\end{wrapfigure}
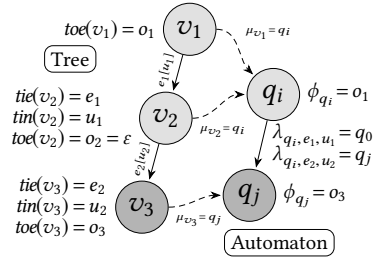

\subsubsection{Positive Scenario Tree Mapping Encoding.}

The goal is to organize a mapping \(\Mapping* : \SetTreeNodes \to \SetStates\) \mbox{between} the nodes of the positive scenario tree~$\PositiveTree$ and the states of the automaton~$\Automaton$.

Variable \(\Mapping{v} \in \SetStates\) denotes the \emph{satisfying state} in which the automaton finishes processing the sequence of scenario elements formed by the path from the root $\TreeRoot$ to the node~$v$.
The root~$\TreeRoot$ itself maps to the initial state: \(\Mapping{\TreeRoot} = \InitialState\).
Passive nodes (those that have $\toe{v} = \varepsilon$) map to the same states as their parents and correspond to the situation, when the automaton ignores an input action, thus: \(
    (\Mapping{v} = q)
    \land
    (\ActualTransitionFunction{q,e,u} = q_0)
\), where $v \in \SetTreeNodesPassive$, $p = \tp{v}$, $q = \Mapping{p}$, $e = \tie{v}$, $u = \tin{v}$.
Active nodes correspond to the situation, when the automaton reacts on an input action by switching the state and producing an output action, which we constrain according to the tree node:
\[
    (\Mapping{v} = q')
    \implies
    (\ActualTransitionFunction{q,e,u} = q')
    \land
    (\OutputEventFunction{q',e,u} = o)
    \land
    \biglandnolim{z \in Z}
    (\AlgorithmFunction{q',e,u,z,b} = b') ,
\]
where $v \in \SetTreeNodesActive$, $p = \tp{v}$, $q = \Mapping{p}$, $q' \in \SetStates$, $e = \tie{v}$, $u = \tin{v}$, $o = \toe{v}$, $z \in Z$, $b = \tov{p,z}$, $b' = \tov{v,z}$.

\subsubsection{\textsc{Basic} Algorithm.}

Constraints declared so far already allow to infer a \emph{computable} automaton that is capable of processing input actions and reacting on them by emitting output actions.
Denote by $\textsc{Basic}(\SetPositiveScenarios, C)$ the procedure of inferring an automaton of size~$C$ satisfying positive scenarios~$\SetPositiveScenarios$. The procedure consists of (1)~building a positive scenario tree, (2)~encoding the automaton structure and the scenario tree mapping, and (3)~delegating to the SAT solver.

\subsubsection{Guard Conditions Structure Encoding.}

In the above reduction, guard conditions were represented in the form of truth tables (by variable~$\TransitionFiring*$), which are not easily human-interpretable, and are not usable in control system development software such as Matlab or nxtSTUDIO~\cite{nxtstudio}, where guard conditions must be explicitly represented with Boolean formulas.
Therefore, we supplement the reduction with an encoding of parse trees of arbitrary Boolean formulas over input variables~$\SetInputVariables$.

Each parse tree is built of~$P$ nodes, where~$P$ is a meta-parameter of the proposed method.
Each node may be either a Boolean operator node or a terminal node representing an input variable.
However, not all formulas require~$P$ nodes, and some nodes may remain unused, \ie not included in the parse tree. We call such nodes \emph{none-typed}.
We define the \emph{size} of a parse tree as the number of \emph{typed} (\ie not none-typed) nodes in~it.
Further in this section we additionally assume that $p \in [1 \dd P]$, $c \in [1 \dd P]$, $x \in \SetInputVariables$, unless stated otherwise.

Variable \(\NodeType{q,k,p} \in \Set{\NodeTypeTerminal, \NodeTypeAnd, \NodeTypeOr, \NodeTypeNot, \NodeTypeNone}\) denotes the type of the $p$-th parse tree node of the guard condition on the $k$-th transition from the state $q$, where $\NodeTypeTerminal$~denotes a~terminal node, \enquote{$\NodeTypeAnd$}, \enquote{$\NodeTypeOr$}, \enquote{$\NodeTypeNot$} \--- logic operators, and $\NodeTypeNone$~denotes a none-typed node.
Variable $\NodeInputVariable{q,k,p} \in\nobreak \SetInputVariables \union\nobreak \Set{0}$ denotes the associated input variable (or its absence). Only terminal nodes have associated input variables: \((\NodeType{q,k,p} = \NodeTypeTerminal) \iff (\NodeInputVariable{q,k,p} \neq 0)\).

Variables $\NodeParent{q,k,p} \in [0 \dd (p\!-\!1)]$ and $\NodeChild{q,k,p} \in \Set{0} \union [(p\!+\!1) \dd P]$ denote, respectively, the parent and the (left) child of the $p$-th node (or their absence, \eg $\NodeParent{q,k,p} = 0$).
These variables are related as follows: \((\NodeChild{q,k,p} = ch) \implies (\NodeParent{q,k,ch} = p)\).
Only typed nodes, except the root ($p = 1$), have parents: \((\NodeParent{q,k,p} \neq 0) \iff (\NodeType{q,k,p} \neq \NodeTypeNone)\).
We do not encode the right child explicitly, but for binary operators we assume that it follows the left one:
\[
    \NodeType{q,k,p} \in \Set{\NodeTypeAnd, \NodeTypeOr}
    \land
    (\NodeChild{q,k,p} = c)
    \implies
    (\NodeParent{q,k,c+1} = p) .
\]

Since each binary operator node must have two children, the $P$-th and $(P\!-\!1)$-th nodes cannot be of type \enquote{$\NodeTypeAnd$} or \enquote{$\NodeTypeOr$}. Similarly, the $P$-th node cannot be of type \enquote{$\NodeTypeNot$}.

Variable $\NodeValue{q,k,p,u} \in \Bool$ denotes the Boolean value of the subformula (represented by the subtree rooted in node $p$) on input $u$. Variable $\TransitionFiring{q,k,u}$ defined earlier is a shortcut for the root node value: \(\TransitionFiring{q,k,u} \iff \NodeValue{q,k,1,u}\).
Terminals have values from the associated input variables; values of non-terminal nodes are calculated according to their types and children values; and none-typed nodes have \texttt{False} values:
\begin{align*}
    (\NodeType{q,k,p} = \NodeTypeTerminal) \land (\NodeInputVariable{q,k,p} = x)
    &\implies
    \biglandnolim{u \in \SetTreeInputs}
    \left[
        \NodeValue{q,k,p,u}
        \iff
        u_x
    \right] ;
\\
    (\NodeType{q,k,p} = \NodeTypeAnd) \land (\NodeChild{q,k,p} = c)
    &\implies
    \biglandnolim{u \in \SetTreeInputs}
    \left[
        \NodeValue{q,k,p,u} \iff \NodeValue{q,k,c,u} \land \NodeValue{q,k,c+1,u}
    \right] ;
\\
    (\NodeType{q,k,p} = \NodeTypeOr) \land (\NodeChild{q,k,p} = c)
    &\implies
    \biglandnolim{u \in \SetTreeInputs}
    \left[
        \NodeValue{q,k,p,u} \iff \NodeValue{q,k,c,u} \lor \NodeValue{q,k,c+1,u}
    \right] ;
\\
    (\NodeType{q,k,p} = \NodeTypeNot) \land (\NodeChild{q,k,p} = c)
    &\implies
    \biglandnolim{u \in \SetTreeInputs}
    \left[
        \NodeValue{q,k,p,u} \iff \neg\NodeValue{q,k,c,u}
    \right] ;
\\
    (\NodeType{q,k,p} = \NodeTypeNone)
    &\implies
    \biglandnolim{u \in \SetTreeInputs}
    \left[
        \neg\NodeValue{q,k,p,u}
    \right] .
\end{align*}

Additionally, we declare auxiliary symmetry-breaking constraints, which force parse tree nodes to be enumerated in BFS order.
Essentially, they are almost identical to BFS constraints for automaton states, but declared for each parse tree separately (for each $q \in \SetStates$, $k \in [1 \dd K]$).
Variable $t^\text{bfs}_{i,j} \in \Bool$ ($1 \leq i < j \leq P$) indicates the existence of a transition from the $i$-th to the $j$-th node: \(
    t^\text{bfs}_{i,j}
    \iff
    (\NodeParent{q,k,j} = i)
\).
Variable $p^\text{bfs}_{j} \in [1 \dd (j\!-\!1)]$ ($j \in [2 \dd P]$) denotes the parent of the $j$-th in the BFS traverse tree: \(
    (p^\text{bfs}_{j} = i)
    \iff
    t^\text{bfs}_{i,j}
    \land
    \biglandnolim{r < i}
    (\neg t^\text{bfs}_{r,j})
\).
Actual BFS constraint is defined as follows: \(
    (p^\text{bfs}_{j} = i)
    \implies
    \biglandnolim{r < i}
    (p^\text{bfs}_{j+1} \neq r)
\).

\subsubsection{Bounding the Guard Conditions Size.}

Additionally, we declare an upper bound for the total size of all guard conditions, \ie the total number of typed parse tree nodes $N$, by imposing a \emph{cardinality constraint}
\(\bigsumnolim{q \in \SetStates, k \in [1 \dd K], p \in [1 \dd P]} \textsc{bool2int}(\NodeType{q,k,p} \neq \NodeTypeNone) \leq N\).

In order to encode this relation in SAT, we use a technique from~\cite{sat-cardinality}. Briefly, this technique consists in declaring a \emph{totalizer}, which encodes the sum in unary form, and a \emph{comparator}, which bounds this sum. For brevity, we omit a formal definition of resulting constraints which can be found in~\cite{sat-cardinality}.

\subsubsection{\textsc{Extended} Algorithm.}

Denote by $\textsc{Extended}(\SetPositiveScenarios, C, P, N)$ the procedure for inferring an automaton which satisfies positive scenarios $\SetPositiveScenarios$, has $C$ states, $P$ nodes in each guard condition parse tree and at most $N$ total nodes in all parse trees.
The procedure consists of (1) building a positive scenario tree, (2) declaring constraints encoding the automaton structure, the scenario tree mapping, the guard conditions structure, and, if parameter~$N$ is specified, a totalizer and a comparator encoding the relation \enquote{total size of guard conditions is less than or equal to $N$}, and (3) delegating to the SAT solver.

\subsubsection{Negative Scenario Tree Mapping Encoding.}

Mapping $\NegativeMapping* : \SetNegativeTreeNodes \to \SetStatesAux$ for the negative scenario tree is similar to the positive one. The key difference is that the negative tree may represent the behavior which the automaton does not have. Moreover, it contains a \emph{looping} behavior, which the automaton is explicitly prohibited to have.

Variable $\NegativeMapping{\negv} \in \SetStatesAux$ denotes the satisfying state (or its absence) of the negative tree node $\negv \in \SetNegativeTreeNodes$, where $\NegativeMapping{\negv} = q_0$ denotes the absence of satisfying state and corresponds to the situation, when the automaton does not have the behavior represented by negative tree.
The root $\NegativeTreeRoot$ maps to the initial automaton state: \(\NegativeMapping{\NegativeTreeRoot} = \InitialState\).
Passive nodes either map to the same states as their parents, or do not map to any state at all (or rather, \emph{map to} $q_0$): \(
    (\NegativeMapping{\negv} = q)
    \lor
    (\NegativeMapping{\negv} = q_0)
\), where $\negv \in \SetNegativeTreeNodesPassive$, $\Negative{p} = \negtp{\negv}$, $q = \NegativeMapping{\Negative{p}}$.
The first case corresponds to the situation, when the automaton ignores an input action: \(
    (\NegativeMapping{\negv} = q)
    \implies
    (\NegativeActualTransitionFunction{q,e,u} = q_0)
\), where $\negv \in \SetNegativeTreeNodesPassive$, $q \in \SetStates$, $e = \negtie{\negv}$, $u = \negtin{\negv}$.
Similarly, active nodes either map to the state in which the automaton goes upon processing an input action, or stay unmapped:
\[
    (\NegativeMapping{\negv} = q')
    \iff
    (\NegativeActualTransitionFunction{q,e,u} = q')
    \land
    (\NegativeOutputEventFunction{q',e,u} = o)
    \land
    \biglandnolim{z \in Z}
    (\NegativeAlgorithmFunction{q',e,u,z,b} = b') ,
\]
where $\negv \in \SetNegativeTreeNodesActive$, $\Negative{p} = \negtp{\negv}$, $q = \Mapping{\Negative{p}}$, $q' \in \SetStates$, $e = \negtie{\negv}$, $u = \negtin{\negv}$, $o = \negtoe{\negv}$, $z \in Z$, $b = \negtov{\Negative{p},z}$, $b' = \negtov{\negv,z}$.
Note that this constraint requires `iff' in contrast to the positive one (where `implication' is enough), because the codomain of $\NegativeMapping*$ is now $\SetStatesAux$, but the constraint is only defined for $q' \in \SetStates$.
Additionally, if some node $\negv \in \SetNegativeTreeNodes$ does not map to any state, then this propagates down the tree: \(
    (\NegativeMapping{\negtp{\negv}} = q_0)
    \implies
    (\NegativeMapping{\negv} = q_0)
\).
Lastly, in order to prohibit the undesired looping behavior represented by back edges, we ensure that the start and the end of each loop either map to different states, or both are unmapped: \(
    \biglandnolim{\negv' \in \negtbe{\negv}}
    \left[
        (\NegativeMapping{\negv} \neq \NegativeMapping{\negv'})
        \lor
        (\NegativeMapping{\negv} = \NegativeMapping{\negv'} = q_0)
    \right]
\), where $\negv \in \SetNegativeTreeNodes$.

\subsubsection{\textsc{Complete} Algorithm.}

Let us denote by $\textsc{Complete}(\SetPositiveScenarios, \SetNegativeScenarios, C, P, N)$ the procedure for inferring an automaton which satisfies positive scenarios $\SetPositiveScenarios$, does not satisfy negative scenarios $\SetNegativeScenarios$, and has $C$ states, $P$ nodes in each guard condition parse tree and at most $N$ total nodes in all parse trees.
The procedure consists of (1)~building both positive and negative scenario trees, (2)~declaring all described constraints, including cardinality constraints if parameter $N$ is specified, and (3)~delegating to the SAT solver.

\subsection{Minimal Model Inference}%
\label{sub:minimal-model-inference}

Proposed methods require the automaton parameters ($C$, $P$ and $N$) to be known \emph{in advance}.
To \emph{automate} the inference of minimal models we use an iterative approach.

\begin{adjustbox}{valign=T, minipage=0.57\textwidth}

\subsubsection{\textsc{Basic-min} Algorithm.}

In order to \emph{quickly} estimate the minimal number of states, we use the $\textsc{Basic}(\SetPositiveScenarios, C)$ algorithm by iterating $C$ starting from 1 until we find a solution \--- satisfying automaton $\Automaton$ with $C_\text{min}$ states.
Let us denote this process as $\textsc{Basic-min}(\SetPositiveScenarios)$ (\cref{algo:basic-min}).

\end{adjustbox}%
\hfill%
\begin{adjustbox}{valign=T, width=0.4\textwidth, raise={4pt}{\height}}
\begin{adjustbox}{width=\linewidth, minipage=0.49\textwidth}
\begin{algorithm}[H]
    \caption{Basic-min($\SetPositiveScenarios$)}%
    \label{algo:basic-min}
    \DontPrintSemicolon

    \KwIn{positive scenarios $\SetPositiveScenarios$}
    \KwOut{automaton $\Automaton$ with minimal number of states $C_\text{min}$}

    \For{$C_{\textit{min}} = 1$ \KwTo $\infty$}{%
        $\Automaton \leftarrow \FnBasic{\SetPositiveScenarios, C_\textit{min}}$\;
        \lIf{$\Automaton \neq \texttt{null}$}{%
            \KwRet{$\Automaton$}
        }
    }

\end{algorithm}%
\end{adjustbox}%
\end{adjustbox}

\begin{wrapfigure}{R}{0.464\textwidth}
    \vspace{-\intextsep}
    \vspace{-2pt}
    \subfile{tex/algo-extended-min}%
    \vspace{-\intextsep}
    \vspace{-10pt}
\end{wrapfigure}

\subsubsection{\textsc{Extended-min} Algorithm.}

Assuming that parameter $P$ is known and $C$ is estimated using the \textsc{Basic-min} algorithm, we minimize the automaton in terms of $N$ as follows.
We declare an upper bound for the total number of parse tree nodes~$N$ and use the \textsc{Extended} algorithm, decreasing~$N$ successively until there is no smaller solution.
The last inferred automaton has $C_\text{min}$ states and its guard conditions have $N_\text{min}$ parse tree nodes in total.
Let us denote this process as $\textsc{Extended-min}(\SetPositiveScenarios, P)$ (\cref{algo:extended-min}).

\subsubsection{\textsc{Extended-min-UB} Algorithm.}

Ultimately, an \emph{automatic} way of determining an appropriate value of parameter $P$ is desirable.
The solution exists when $P$ is large enough to capture the necessary guard conditions complexity.
The simplest strategy is to iterate $P$ starting from 1 and use $\textsc{Extended-min}(\SetPositiveScenarios, P)$ until we find a solution \--- automaton with $N = N^{*}_\text{min}$ \--- for some $P^{*}$.
However, there may exist some value $P' > P^{*}$ for which the corresponding $N'_\text{min}$ is even smaller than $N^{*}_\text{min}$. Therefore, in order to obtain the globally minimal automaton in terms of $N$, we shall continue the search process for $P > P^{*}$ up to a theoretical upper bound as described in
\ifappendices{%
    \Cref{app:automatic-p-upper-bound}%
}{%
    \appendixname~A~\cite{fbSAT-full}%
}%
, where we define the $\textsc{Extended-min-UB}(\SetPositiveScenarios, w)$ algorithm, which allows to \emph{automatically} infer the minimal automaton in terms of $C$, $P$ and~$N$ from the given positive scenarios~$\SetPositiveScenarios$.
Parameter $w \geq 0$ is a heuristic threshold plateau width. When $w = 0$, the algorithm is equivalent to the simplest strategy of searching~$P$ until the first SAT\@. When $w = \infty$, the algorithm continues to iterate~$P$ until an upper bound, resulting in the globally minimal~$N_\text{min}$. Other values enable a heuristic providing a trade-off between minimality and execution time.

\subsection{Counterexample-Guided Inductive Synthesis}

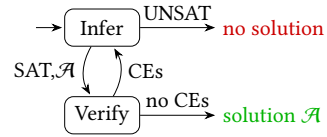
\begin{wrapfigure}{R}{0.35\textwidth}
    \vspace{-\intextsep}
    \centering
    \begin{adjustbox}{width=\linewidth}
        \begin{tikzpicture}[
    >={Stealth[]},
    shorten >=1pt,
    auto,
    initial text={},
    every state/.style={
        rectangle,
        rounded corners,
        inner sep=0pt,
        minimum height=6mm,
        minimum width=12mm,
        align=center,
    },
]
    \def\HGap{12mm}
    \def\VGap{8mm}
    \colorlet{good}{green!70!black}
    \colorlet{bad}{red!80!black}

    \node[initial,state] (infer)                              {Infer};
    \node[state]         (verify)     [below=\VGap of infer]  {Verify};
    \node[text=bad]      (nosolution) [right=\HGap of infer]  {no solution};
    \node[text=good]     (solution)   [right=\HGap of verify] {solution $\mathcal{A}$};

    \path[->]
        (infer)  edge[bend right,left]     node{SAT,$\mathcal{A}$} (verify)
        (verify) edge[bend right,right]    node{CEs}               (infer)
        (infer)  edge[above,shorten >=0pt] node{UNSAT}             (nosolution)
        (verify) edge[above,shorten >=0pt] node{no CEs}            (solution)
    ;
\end{tikzpicture}%
%
    \end{adjustbox}
    \vskip-\abovecaptionskip
    \vspace{2pt}
    \caption{CEGIS loop}%
    \label{fig:cegis-approach}
    \vskip-\belowcaptionskip
\end{wrapfigure}

In order to make the inferred automaton not only satisfy given positive scenarios, but also comply with an LTL specification, we use a counterexample-guided inductive synthesis (CEGIS)~\cite{solar-lezama-2006} iterative approach.
Each CEGIS iteration consists of inferring an automaton~$\Automaton$, verifying an LTL specification~$\LTLSpec$ using a model checker, and supplementing the negative scenario tree with obtained counterexamples, if any. The process shown in Fig.~\ref{fig:cegis-approach} repeats until there are no more counterexamples, thus, the inferred automaton complies with the given LTL specification~$\LTLSpec$.
Denote by $\textsc{Complete-CEGIS}(\SetPositiveScenarios, \LTLSpec, C, P, N)$ the procedure implementing the CEGIS, where arguments are similar to the one in the $\textsc{Complete}$ algorithm.
Also, denote by $\textsc{Complete*-CEGIS}(\SetPositiveScenarios, \LTLSpec, w)$ the procedure consisting of a call to $\textsc{Extended-min-UB}(\SetPositiveScenarios, w)$ followed by a $\textsc{Complete-CEGIS}(\SetPositiveScenarios, \LTLSpec, C^{*}, P^{*}, N=\infty)$ with estimated $C^{*}$ and $P^{*}$ \--- note that $N$ is unbounded.

\subsubsection{\textsc{Complete*-min-CEGIS} Algorithm.}

Consider an automaton $\Automaton$ produced by the \textsc{Complete-CEGIS} algorithm. If we start minimizing the total size of guard conditions~$N$, the automaton will most likely stop complying with the LTL specification, though the already obtained negative scenarios will still not be satisfied.
Therefore, we propose to maintain a minimal model on each CEGIS iteration.
We begin with a model produced by $\textsc{Extended-min-UB}(\SetPositiveScenarios, w)$ and continue by starting a CEGIS loop via $\textsc{Complete-CEGIS}(\SetPositiveScenarios, \LTLSpec, C^{*}, P^{*}, N^{*})$ with estimated $C^{*}$, $P^{*}$ and $N^{*}$. The UNSAT result indicates that $N^{*}$ is too small for an automaton to comply with the given LTL specification $\LTLSpec$, hence we increase it and continue the CEGIS\@. Note that this is the only moment we stop solving incrementally, because we weaken the contraints (upper bound for $N$).
Let us denote the described process as $\textsc{Complete*-min-CEGIS}(\SetPositiveScenarios, \LTLSpec, w)$.

\subsection{The \textsc{fbSAT} Tool}%
\label{sec:fbsat}

We implemented the proposed methods in a command-line tool \textsc{fbSAT}~\cite{fbSAT-tool} (\url{www.github.com/ctlab/fbSAT}) written in Kotlin.
\textsc{fbSAT} takes as input the execution scenarios and the parameters necessary for the specified method, and infers an automaton satisfying given scenarios and LTL properties.
As~a~backend, \textsc{fbSAT} is able to use any SAT solver.
In our work we use the \texttt{CryptoMiniSat}~\cite{cryptominisat} SAT solver through the \texttt{incremental-cryptominisat}~\cite{incremental-cryptominisat} wrapper, utilizing its ability in \emph{incremental} SAT solving, greatly descreasing total solving time, as our minimization problems are inherently incremental.

\section{Case study: Pick-and-Place manipulator}%
\label{sec:experimental-evaluation}

The experimental evaluation of proposed methods was done on a case study devoted to the inference of a finite-state model of the controller for a Pick-and-Place (PnP) manipulator~\cite{patil-pnp} shown in \cref{fig:pnp-manipulator}.
We also performed an evaluation on random automata
\ifappendices{%
    (\Cref{app:case-study-random}).%
}{%
    (\appendixname~B~\cite{fbSAT-full}).%
}
Experiments were conducted on a computer with an Intel(R) Core\texttrademark{} i5-7200U CPU @ 2.50~GHz and 8~GB of RAM\@.

\begin{wrapfigure}{R}{0.36\textwidth}
    \vspace{-\intextsep}
    \centering
    \includegraphics[width=\linewidth]{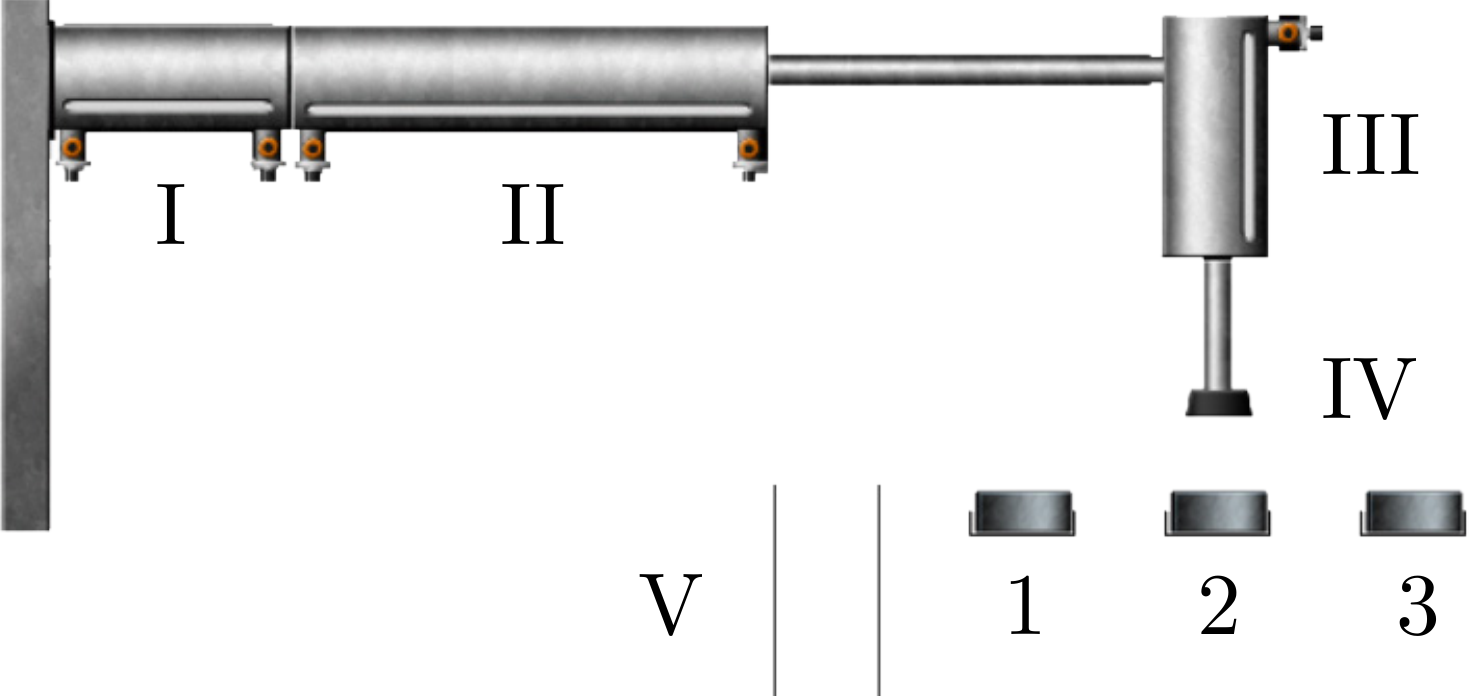}
    \vskip-\abovecaptionskip
    \vspace{2pt}
    \caption{Pick-and-Place manipulator}%
    \label{fig:pnp-manipulator}
    \vskip-\belowcaptionskip
\end{wrapfigure}

The PnP manipulator consists of two horizontal pneumatic cylinders (I,~II), one vertical cylinder (III), and a suction unit (IV) for picking up work pieces. When a work piece appears on one of the input sliders (1, 2, 3), the horizontal cylinders position the suction unit on top of the work piece, the vertical cylinder lowers the suction unit where it picks up the work piece and then moves in to the output slider (V). The control system is implemented using IEC~61499 FBs in nxtSTUDIO~\cite{nxtstudio}. The controller is a basic FB with 10 input and 7 output variables.
The controller of the PnP system uses the following signals from the plant represented by Boolean input variables:
\begin{itemize}[nosep]
    \item \com{c1Home}/\com{c1End} \--- is horizontal cylinder I in fully retracted/extended position;
    \item \com{c2Home}/\com{c2End} \--- is horizontal cylinder II in fully retracted/extended position;
    \item \com{vcHome}/\com{vcEnd} \--- is vertical cylinder III in fully retracted/extended position;
    \item \com{pp1}/\com{pp2}/\com{pp3} \--- is a WP present on input slider 1/2/3;
    \item \com{vac} \--- is the vacuum unit IV on.
\end{itemize}
\vspace{1pt}
The following commands can be issued by the controller to the plant:
\begin{itemize}[nosep]
    \item \com{c1Extend}/\com{c1Retract} \--- extend/retract cylinder I;
    \item \com{c2Extend}/\com{c2Retract} \--- extend/retract cylinder II;
    \item \com{vcExtend} \--- extend cylinder III;
    \item \com{vacuum\_on}/\com{vacuum\_off} \--- turn the vacuum unit on/off.
\end{itemize}

The purpose of this case study was to infer a finite-state model of this controller~FB\@.
The process of capturing scenarios for the Pick-and-Place manipulator controller is described in~\cite{fbCSP}. We used sets of scenarios of various sizes: 1,~10, 39 and 49 scenarios in each.

\subsubsection{Inference of automata with minimal guard conditions from positive scenarios.}

In the first set of experiments we compare methods that infer models from positive scenarios with explicit regard of guard conditions size.
Our method was compared to the two-stage approach from~\cite{chivilikhin-19}, where on the first stage a basic automaton model is inferred with a SAT solver, and then this model's guard conditions are minimized with a CSP solver \wrt given scenarios.
Note that the two-stage method has already been shown in~\cite{chivilikhin-19} to be superior to EFSM-tools~\cite{efsm-tools}.

We apply the proposed \textsc{Extended-min-UB} method to infer an automaton with the minimal number of states~$C_\text{min}$ and total size of guard conditions~$N_\text{min}$. Three values of the $w$ parameter were used: $w = 0$ for the case when first solution found is considered final, $w = 2$ for the case with the proposed heuristic applied, and $w=\infty$ for the \enquote{without heuristic} case.
Results are summarized in \cref{tab:experimental-results-pnp}, where for the two-stage method from~\cite{chivilikhin-19}: $C_\text{min}$ \--- minimal number of states, $T_\text{min}$ \--- minimal number of transitions, $N_\text{sum}$ \--- size of guard conditions; and for \textsc{Extended-min-UB}: $w$ \--- maximum width  of local minima plateau, $P$ \--- maximum guard condition size, $T$ \--- number of transitions, $N_\text{min}$ \--- minimal total size of guard conditions.
Results indicate that \textsc{Extended-min-UB} produces compact automata: in studied cases, using $w=2$ already yields the optimal result in terms of $N_\text{min}$.

\begin{table}[!htb]
    \centering
    \vspace{4pt}  %
    \vskip-\abovecaptionskip
    \caption{Inference of automata with minimal guard conditions from positive scenarios}%
    \label{tab:experimental-results-pnp}
    \vskip-\belowcaptionskip
    \vspace{2pt}
\begin{adjustbox}{max width=0.95\textwidth, max totalheight=4cm}
\begin{tabular}{
    cc
    @{\hspace{6pt}}
    cccc
    @{\hspace{6pt}}
    cccc
    @{\hspace{6pt}}
    cccc
    @{\hspace{6pt}}
    cccc
}
\toprule
    \multirowcell{3}{$\SetPositiveScenarios$} & \multirowcell{3}{$|\PositiveTree|$}
    & \multicolumn{4}{c}{\multirowcell{2}{Two-stage~\cite{chivilikhin-19}}}
    & \multicolumn{12}{c}{\textsc{Extended-min-UB}}
\\ & & & & &  %
    & \multicolumn{4}{c}{$(w = 0)$}
    & \multicolumn{4}{c}{$(w = 2)$}
    & \multicolumn{4}{c}{$(w = \infty)$}
\\ &  %
    & time,\,s. & $C_{\min}$ & $T_{\min}$ & $N_\text{sum}$
    & time,\,s. & $P$ & $T$ & $N_{\min}$
    & time,\,s. & $P$ & $T$ & $N_{\min}$
    & time,\,s. & $P$ & $T$ & $N_{\min}$
\\\cmidrule(lr){1-2} \cmidrule(lr){3-6} \cmidrule(lr){7-10} \cmidrule(lr){11-14} \cmidrule(lr){15-18}
    $\SetScenarios^{(1)}$ & 24
    & 8 & 6 & 8 & 15  %
    & 3 & 3 & 8 & 14  %
    & 4 & 3 & 8 & 14  %
    & 5 & 3 & 8 & 14  %
\\ %
    $\SetScenarios^{(10)}$ & 234
    &   3 & 8 & 17 & 36  %
    &  17 & 3 & 18 & 38  %
    &  58 & 5 & 16 & 25  %
    &  87 & 5 & 16 & 25  %
\\ %
    $\SetScenarios^{(39)}$ & 960
    &  13 & 8 & 15 & 32  %
    &  41 & 3 & 18 & 38  %
    & 124 & 5 & 16 & 25  %
    & 162 & 5 & 16 & 25  %
\\ %
    $\SetScenarios^{(49)}$ & 2939
    &    36 & 8 & 18 & 60  %
    &   602 & 5 & 18 & 44  %
    &  4305 & 6 & 16 & 39  %
    & 51666 & 6 & 16 & 39  %
\\\bottomrule
\end{tabular}
\end{adjustbox}%
\end{table}%

\subsubsection{Comparison with LTL synthesis tools.}

We considered LTL synthesis tools BoSy~\cite{bosy} and G4LTL-ST~\cite{g4ltl-st}, which
accept LTL specifications as input.
Comparison was only done for synthesis from scenarios, which were converted to LTL formulas.
For BoSy we considered a simplified version of scenario $\SetScenarios^{(1)}$, for which passive elements were removed, leaving only 8 scenario elements.
The input-symbolic version of BoSy was the only one that worked for this example, generating a solution with 9 states and 17 transitions in 273~sec.
For G4LTL-ST we selected the number of unroll steps (mandatory parameter of this tool) equal to the length of the largest scenario.
For $\SetScenarios^{(1)}$ a solution with 10 states (though with verbose guard conditions) was found in 10~sec.
Larger sets of scenarios required 16 unroll steps, and runs failed with a memory limit of 8~GB\@.
As expected, experiments showed that LTL synthesis tools are not well-suited for inference of models from finite-length scenarios.
Experiments with LTL properties were not considered due to (1)~poor performance on scenarios, and (2)~lack of support for general\-/form NuSMV plant model, which is crucial for synthesis from liveness properties.

\subsubsection{Inference of automata from positive scenarios and LTL properties.}

The third set of experiments is devoted to CEGIS\@.
In order to enable use of liveness LTL properties, verification of candidate models with NuSMV was performed in a closed loop~\cite{closed-loop} with a manually prepared formal model of the plant \--- the PnP manipulator.
This model defines plant state and its actions implied by controller commands.
The set of considered LTL properties (\Cref{tab:ltl-properties}) includes safety properties $\varphi_1$\--$\varphi_6$ (controller does not lead the system to an unsafe state) and liveness properties $\varphi_7$\--$\varphi_{10}$ (something useful eventually happens).
Properties $\varphi_1$\--$\varphi_7$ are fixed and used in all experiments, while use of $\varphi_8$\--$\varphi_{10}$ varies.
We concentrate on these last properties, which define that whenever a WP is placed on some input slider, it will eventually be removed.
Note that for the original PnP system~\cite{patil-pnp} only $\varphi_8$ is satisfied, and $\varphi_9$\--$\varphi_{10}$ are false (the controller is not wait-free for sliders 2 and 3 \--- if a WP is always present on slider 1, WPs from sliders 2 and 3 will never be picked up).
Therefore, we consider the property for each input slider separately, assuming that WPs never appear on other input sliders.
For the experiment with $\varphi_{9}$ we use a special set of scenarios $\SetScenarios^{(1)\prime\prime}$, which consists only of the second scenario describing a single processing of a WP from slider 2.
Similarly, for $\varphi_{10}$ the set $\SetScenarios^{(1)\prime\prime\prime}$ is used that describes one processing of a WP from slider 3.

\begin{table}[!tb]
    \centering
    \vskip-\abovecaptionskip
    \caption{Temporal properties for the Pick-and-Place system}%
    \label{tab:ltl-properties}
    \vskip-\belowcaptionskip
    \vspace{2pt}
\begin{adjustbox}{max width=\linewidth}
\begin{tabular}{ L m{6.3cm} @{\hspace{4pt}} | @{\hspace{4pt}} m{8cm} }
\toprule
    &
    \thead{Property} &
    \thead{Description}
\\\midrule
    \multicolumn{3}{c}{Fixed part}
\\\midrule
    \varphi_1 &
    $\temp{G}(\neg(\com{c1Extend} \land \com{c1Retract}))$ &
    Cylinder I must not be issued commands to extend and retract simultaneously.
\\
    \varphi_2 &
    $\temp{G}(\neg(\com{c2Extend} \land \com{c2Retract}))$ &
    Similar property for cylinder II.
\\
    \varphi_3 &
    $\temp{G}(\neg(\com{vacuum\_on} \land \com{vacuum\_off}))$ &
    Similar property for the vacuum unit.
\\
    \varphi_4 &
    $\temp{G}(\neg\com{vcHome} \land \neg\com{vcEnd} \rightarrow \com{c1Home} \lor \com{c1End})$ &
    If the vertical cylinder is in the intermediate position, cylinder I must be either in home or end position.
\\
    \varphi_5 &
    $\temp{G}(\neg\com{c1Home} \land \neg\com{c1End} \rightarrow \com{vcHome} \lor \com{vcEnd})$ &
    If cylinder I is in the intermediate position, the vertical cylinder must be either in home or end position.
\\
    \varphi_6 &
    $\temp{G}(\com{all\_home} \land \neg\com{pp1} \land \neg\com{pp2} \land \neg\com{pp3} \land \neg\com{lifted} \rightarrow \temp{X}(\neg\com{c1Extend} \land \neg\com{c2Extend} \land \neg\com{vcExtend}))$ &
    If all cylinders are in home position and no WP should be processed, no commands to move any cylinders should be issued.
\\
    \varphi_7 &
    $\temp{G}(\com{lifted} \rightarrow \temp{F}(\com{dropped}))$ &
    If a WP is lifted from the input slider it must eventually be dropped to the output slider.
\\\midrule
    \multicolumn{3}{c}{Variable part (one at a time)}
\\\midrule
    \varphi_8 &
    $\temp{G}(\com{pp1} \rightarrow \temp{F}(\com{vp1}))$ &
    If a WP appears on input slider 1 it must be eventually lifted.
\\
    \varphi_9 &
    $\temp{G}(\com{pp2} \rightarrow \temp{F}(\com{vp2}))$ &
    If a WP appears on input slider 2, it must be eventually lifted.
\\
    \varphi_{10} &
    $\temp{G}(\com{pp3} \rightarrow \temp{F}(\com{vp3}))$ &
    If a WP appears on input slider 3, it must be eventually lifted.
\\\bottomrule
\end{tabular}%
\end{adjustbox}%
\end{table}%

Three algorithms are compared: the proposed \textsc{Complete*-min-CEGIS}, \textsc{Complete*-CEGIS}, and the CEGIS-extension of \textsc{fbCSP}~\cite{chivilikhin-18}.
EFSM-tools~\cite{efsm-tools}, BoSy~\cite{bosy} and G4LTL-ST~\cite{g4ltl-st} were not considered here, first of all due to poor performance on scenarios only.
For both our algorithms we use $w = 2$, as this value has shown a good performance in our initial study.
Apart from running time and $N$, we measure $N_{\text{init}}$ (for the automaton built from positive scenarios using \textsc{Extended-min-UB}) and the number of CEGIS-iterations (\#iter).
The inferred models were tested in nxtSTUDIO \--- loaded into the simulation environment and checked for compliance with the desired behavior.
Experimental results are summarized in Table~\ref{tab:ltl-experiment}.
\ifappendices{%
    An example of automata generated using \textsc{Complete*-min-CEGIS} algorithm from scenarios $\SetScenarios^{(1)}$ and $\SetScenarios^{(39)}$ and LTL specification $\varphi_1$\==$\varphi_8$ are shown in ~\cref{fig:example-automaton-complete-min-cegis-tests-1} and~\cref{fig:example-automaton-complete-min-cegis-tests-39} (see Appendix).%
}{%
    An example of an automaton generated using \textsc{Complete*-min-CEGIS} algorithm from scenarios $\SetScenarios^{(1)}$ and LTL specification $\varphi_1$\==$\varphi_8$ is shown in~\cref{fig:example-automaton-complete-min-cegis-tests-1}.%
}

\begin{table}[!htb]
    \centering
    \vskip-\abovecaptionskip
    \vspace{4pt} %
    \caption{Results of CEGIS experiments for PnP controller inference}%
    \label{tab:ltl-experiment}
    \vskip-\belowcaptionskip
    \vspace{2pt}
\begin{adjustbox}{max width=0.95\textwidth, max totalheight=5cm}
\begin{tabular}{
    ccc
    @{\hspace{8pt}}
    c @{\hspace{4pt}} c @{\hspace{12pt}} cc
    @{\hspace{8pt}}
    ccc
    @{\hspace{8pt}}
    ccc
}
\toprule
    \multirowcell{2}{LTL properties} &
    \multirowcell{2}{Scenarios} &
    \multirowcell{2}{$N_{\text{init}}$} &
    \multicolumn{4}{c}{\textsc{Complete*-min-CEGIS}} &
    \multicolumn{3}{c}{\textsc{Complete*-CEGIS}} &
    \multicolumn{3}{c}{\textsc{fbCSP+LTL}~\cite{chivilikhin-18}}
\\ & &
    & time,\,s. & \#iter & $P$ & $N$  %
    & time,\,s. & \#iter & $N$        %
    & time,\,s. & \#iter & $N$        %
\\\cmidrule(lr){1-3} \cmidrule(lr){4-7} \cmidrule(lr){8-10} \cmidrule(lr){11-13}
    \multirowcell{3}{$\temp{G}(\mbox{\com{pp1}} \rightarrow \temp{F}(\mbox{\com{vp1}}))$}
    & $\SetScenarios^{(1)}$ & 14
    & 55 & 273 & 3 & 16  %
    & 19 & 79 & 33       %
    & >12h & >500 & --   %
\\
    & $\SetScenarios^{(10)}$ & 25 &
    242 & 20 & 5 & 28 &  %
    60 & 2 & 118 &       %
    613 & 10 & 40        %
\\
    & $\SetScenarios^{(39)}$ & 25 &
    425 & 70 & 5 & 28 &  %
    142 & 20 & 128 &     %
    1019 & 2 & 41        %
\\\cmidrule(lr){1-3} \cmidrule(lr){4-7} \cmidrule(lr){8-10} \cmidrule(lr){11-13}
    \multirowcell{1}{$\temp{G}(\mbox{\com{pp2}} \rightarrow \temp{F}(\mbox{\com{vp2}}))$} & $\SetScenarios^{(1)\prime\prime}$ & 14 &
    63 & 193 & 3 & 16 &  %
    37 & 188 & 29 &      %
    >12h & >500 & --     %
\\\cmidrule(lr){1-3} \cmidrule(lr){4-7} \cmidrule(lr){8-10} \cmidrule(lr){11-13}
    \multirowcell{1}{$\temp{G}(\mbox{\com{pp3}} \rightarrow \temp{F}(\mbox{\com{vp3}}))$} & $\SetScenarios^{(1)\prime\prime\prime}$ & 14 &
    116 & 353 & 3 & 16 &  %
    400 & 1450 & 34 &     %
    >12h & >500 & --      %
\\\bottomrule
\end{tabular}
\end{adjustbox}%
\end{table}%

Solutions found with CEGIS methods are always larger than ones constructed from scenarios only (in terms of $N$).
This indicates that the used sets of scenarios are incomplete and do not covered considered specifications completely.
Then, \textsc{Complete*-min-CEGIS} always finds the smallest solutions and is always faster than~\cite{chivilikhin-18}.
Most interestingly, \textsc{Complete*-min-CEGIS} allows efficiently constructing models for scenarios $\SetScenarios^{(1)}$, $\mathcal{S}^{(1)\prime\prime}$, $\mathcal{S}^{(1)\prime\prime\prime}$ \--- these scenarios do not \enquote{cover} corresponding liveness properties of interest (\eg $\temp{G}(\mbox{\com{pp1}} \rightarrow \temp{F}(\mbox{\com{vp1}}))$) in the sense that the scenario describes only a single processing of a WP\@. The existing method~\cite{chivilikhin-18} failed on these cases, while the proposed approach succeeds with ease.
Lastly, \textsc{Complete*-CEGIS} allows constructing models fast, but loosing the guard conditions minimality.

\begin{figure}[hbt!]
    \centering
    \includegraphics[max width=\linewidth, max height=\dimexpr\pagegoal-\pagetotal-\baselineskip-\abovecaptionskip-\belowcaptionskip-10pt\relax]{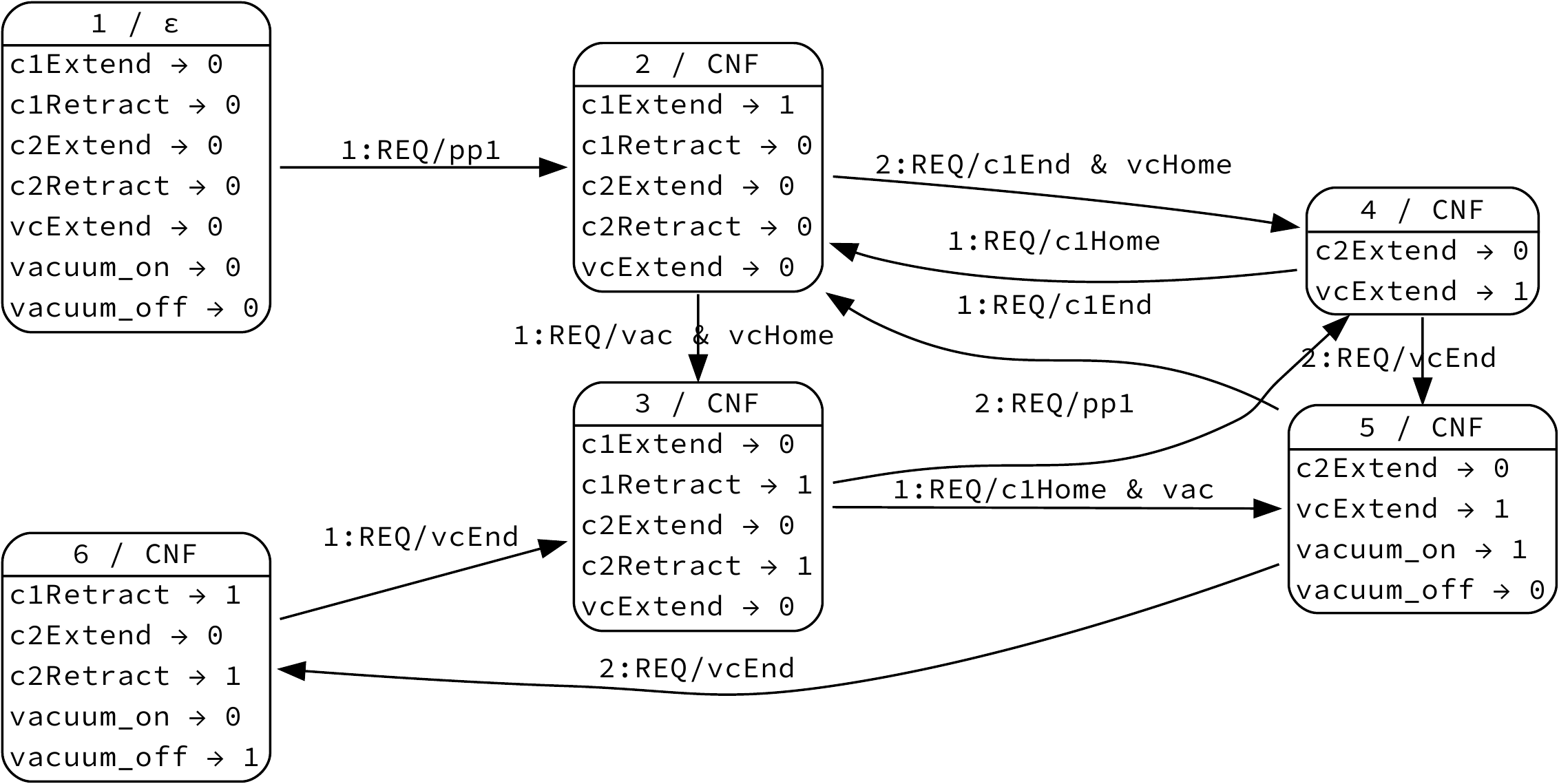}
    \vskip-\abovecaptionskip
    \vspace{4pt}
    \caption{Example automaton generated using \textsc{Complete*-min-CEGIS} algorithm from $\SetScenarios^{(1)}$}%
    \label{fig:example-automaton-complete-min-cegis-tests-1}
    \vskip-\belowcaptionskip
\end{figure}

\section{Discussion}%
\label{sec:discussion}

The proposed methods allow to infer minimal finite-state models of function blocks from a given specification: execution scenarios and LTL properties.
Execution scenarios are commonly derived from existing systems under learn (SULs) or by means of simulation of existing systems' models, and LTL properties are commonly written manually.
Note that synthesized models are, in general, not equivalent to the original SUL, and their behavior may differ in situations not covered by the specification.
We assume that the given specification covers the necessary behavior, and we believe that the generalization through minimization helps producing models that conform with the original one.
Note that it is impossible to formally check neither the conformance of the SUL to the LTL specification, nor the equivalence of the synthesized model and the  SUL: we do not have explicit access to the SUL (\ie to the internal structure and source code) due to purely \enquote{passive learning} problem statement coming from the industry.

\section{Conclusion and Future Work}%
\label{sec:conclusion}

We have proposed a SAT-based approach for inference of minimal FB models from execution scenarios and LTL properties, and implemented it in the tool \textsc{fbSAT}.
The proposed approach is the only one that allows direct minimization of guard conditions complexity of synthesized automata.
In particular, the \textsc{Extended-min-UB} algorithm is guaranteed to find the solution with globally minimal complexity of guard conditions.
Experiments showed that the suggested approach outperforms existing ones and demonstrates predictable scalability on random instances.

Future research may include synthesis of modular automata and applying other encodings for non-Boolean variables and cardinality constraints.
In addition, developed encodings of tree-form arbitrary guard conditions may be used to augment other SAT-based methods for state machine synthesis.
In particular, adding our guard conditions to BoSy~\cite{bosy,not-bosy} may aid in decreasing the sizes of generated transition systems.

\subsubsection*{Acknowledgements.}

This work was funded by the Government of Russian Federation (Grant 08-08).

\biblio

\ifappendices{
    \clearpage
    \begin{subappendices}
    \renewcommand{\setthesection}{\Alph{section}}
    \crefalias{section}{appsection}

\section{Automatic search for best value of \texorpdfstring{$P$}{P}}%
\label{app:automatic-p-upper-bound}

In this appendix we describe in detail the process of searching for the \emph{best} value of $P$ in the sense of minimizing the corresponding $N_\text{min}$ obtained using \textsc{extended-min} algorithm.

Consider ${P = P'}$; ideally, we expect that all guard conditions will be of size 1, and only one of them will be of size $P'$. Also, ideally, there are exactly $T_\text{min}$ guards, therefore, the ideal minimal total size of guard conditions is ${N'_\text{min} = T_\text{min} - 1 + P''}$.
Let us denote by $N^\text{best}_\text{min}$ the best, \ie the most minimal value found so far.
Ultimately, we are looking for $N'_\text{min} < N^\text{best}_\text{min}$, thus ${T_\text{min} - 1 + P' < N^\text{best}_\text{min}}$, from where the upper bound for $P$ is ${P' \leq N^\text{best}_\text{min} - T_\text{min}}$.

The process of searching $P$ up to the upper bound can take an extensive amount of time. Hence, we propose the following heuristic.
Consider the two successive values $P'$ and ${P'' = P' + 1}$, and the corresponding values $N'_\text{min}$ and $N''_\text{min}$. The equality ${N'_\text{min} = N''_\text{min}}$ indicates the local minimum (plateau).
As we go further by incrementing the value of $P''$, the remaining equality extends the plateau width. By choosing the critical plateau width $w$, on which to stop incrementing $P$, we provide a trade-off between the execution time and global minimality of the solution. In practice, an arbitrary choice of ${w = 2}$ showed good performance in our initial studies.
It is worth noting that with this heuristic applied, our proposed method remains \emph{exact} in the sense that inferred automata still satisfy given positive scenarios $\mathcal{S}^{+}$.

Let us denote by $\textsc{extended-min-ub}(\mathcal{S}^{+}, w)$ the minimization process described above. It is depicted by \cref{algo:extended-min-ub} and consists of two stages.
First, we estimate the automaton parameters $C_\text{min}$ and $T_\text{min}$ using $\textsc{basic-min}^{*}$ algorithm.
Note that by $\textsc{basic-min}^{*}$ we denote the algorithm which combines the \textsc{basic-min} and the minimization of $T$ using the same technique as in \textsc{extended-min} for $N$.
Second, we iterate $P$ starting from 1 and use the \textsc{extended-min} algorithm to infer an automaton.
We stop the search in two cases: if current $P$ is greater than the upper bound (${N^\text{best}_\text{min} - T_\text{min}}$), or if current local minumum width is greater than the arbitrary threshold $w$.

\vspace{4pt}
\begin{adjustbox}{width=0.9\textwidth, minipage=1.1\textwidth, center}
\begin{algorithm}[H]
    \caption{Extended-min-UB($\SetPositiveScenarios$, $w$)}%
    \label{algo:extended-min-ub}
    \DontPrintSemicolon

    \KwIn{positive scenarios $\SetPositiveScenarios$,
          maximum plateau width $w$}
    \KwOut{automaton $\Automaton$
           with minimal number of states $C_\text{min}$
           and guard conditions size $N_\text{min}$}

    \BlankLine
    $\Automaton_\textit{basic} \leftarrow \FnBasicMin{\SetPositiveScenarios}$\;
    $T_\textit{min} \leftarrow \FnGetT{\Automaton_\textit{basic}}$\;
    $C_\textit{min} \leftarrow \FnGetC{\Automaton_\textit{basic}}$\;
    $N^\textit{best}_\textit{min} \leftarrow N^\textit{prev}_\textit{min} \leftarrow P_\textit{low} \leftarrow \infty$\;

    \For{$P = 1$ \KwTo $\infty$}{%
        \lIf{$P > (N^\textit{best}_\textit{min} - T_\textit{min})$}{%
            \Break
            \tcp*[h]{upper bound reached}
        }
        \lIf{$(P - P_\textit{low}) > w$}{%
            \Break
            \tcp*[h]{max width reached}
        }

        $\Automaton \leftarrow \FnExtendedMin{\SetPositiveScenarios, C_\textit{min}, P}$\;
        \If{$\Automaton \neq \texttt{null}$}{%
            $N_\textit{min} \leftarrow \FnGetN{\Automaton}$\;
            \lIf{$N_\textit{min} < N^\textit{best}_\textit{min}$}{%
                $N^\textit{best}_\textit{min} \leftarrow\ N_\textit{min}$
                \tcp*[h]{update best found $N$}
            }
            \lIf{$N_\textit{min} \neq N^\textit{prev}_\textit{min}$}{%
                $P_\textit{low} \leftarrow P$
                \tcp*[h]{update local minimum}
            }
            $N^\textit{prev}_\textit{min} \leftarrow N_\textit{min}$\;
        }
    }

    \KwRet{$\Automaton$}
\end{algorithm}%
\end{adjustbox}%

\subfilebiblio

\section{Case Study: Random Automata}
\label{app:case-study-random}

In order to test our tool \textsc{fbSAT} on more instances, we perform an evaluation on randomly generated automata in this case study.

The first step is to generate random automata.
We chose the automaton parameters similar to the parameters of the model inferred in the \enquote{Case Study: PnP Manipulator} (\cref{sec:experimental-evaluation}): number of states $C = 8$, number of transitions $T$ up to $C^2$, one input and one output event, ${|X| = 10}$ input and ${|Z| = 7}$ output variables.
Also, we selected an additional value ${|X| = 5}$ to compare with simpler models.

The second step is to simulate execution scenarios. We start in the initial automaton state and consequently choose a random input event and random input variable values. The automaton reacts on these input actions and produces output actions, forming an execution scenario.
We performed a simulation of two sets of scenarios: (1) 10 scenarios of length 100 each, and (2) 50 scenarios of length 50 each.
Note that this random walk corresponds to a situation when the plant has random dynamics.
Hence, these randomly simulated instances are most likely harder than real-world instances, since real-world plants (such as the PnP manipulator) do not have random dynamics.

The next step is to infer the minimal automaton from the simulated scenarios using the \textsc{extended-min} algorithm.
And the final step is to validate the inferred automaton. We use the \enquote{forward check} validation approach from~\cite{ulyantsev-efsm-sat} consisting in generating a large validation set of scenarios and checking whether the inferred automaton satisfies them. The metric here is the percentage $p$ of satisfied scenarios. We expect to confirm that high coverage of target automata by scenarios leads to good validation results.

\begin{table}[!htb]
    \centering
    \vspace{4pt}
    \vskip-\abovecaptionskip
    \caption{Results for the Random Automata Case Study}%
    \label{tab:results-random}
    \vskip-\belowcaptionskip
    \vspace{2pt}
    \renewcommand{\arraystretch}{0.95}
    \setlength{\tabcolsep}{6pt}
\begin{tabular}{CCCCCCC}
    \toprule
    \thead{|\mathcal{S}|} & \thead{|s|} & \thead{|X|} & \thead{\bar{t} \pm \sigma} & \thead{\bar{p}, \%} & \thead{100\%p} & \\
    \midrule
    10 & 100 &  5 & 48 \pm 38 & 72 &  7 \text{ of } 30 \\
    10 & 100 & 10 & 535 \pm 692 & 30 &  0 \text{ of } 31 \\
    50 &  50 &  5 & 148 \pm 54 & 98 & 41 \text{ of } 56 \\
    50 &  50 & 10 & 991 \pm 702 & 95 &  5 \text{ of } 10 \\
    \bottomrule
\end{tabular}
\end{table}

The results are presented in \cref{tab:results-random}, where $|\mathcal{S}|$ is the number of scenarios, $|s|$~-- length of each scenario, $|X|$~-- number of input variables, $\bar{t}$~-- mean solving time (in seconds), $\sigma$~-- standard deviation, $\bar{p}$~-- mean \enquote{forward check} validation percentage, $\text{100\%p}$~-- number of instances with 100\% validation.
Additionally, the results are shown in \cref{fig:results-random}, where on the left the distribution of execution time is shown, and on the right -- the distrubution of \enquote{forward check} percentage. Both plots are grouped by number of scenarios $|\mathcal{S}|$ and number of input variables $|X|$.

Experimental results (\cref{fig:results-random}) indicate that with sufficient coverage of the target automaton by execution scenarios (in this case, 50 scenarios of length 50 each), our approach allows to identify the exact behavior of the automaton with high probability.
However, inference from large sets of scenarios requires sufficient computational resources.

\begin{figure}[!htb]
    \centering
    \includegraphics[max width=0.9\textwidth]{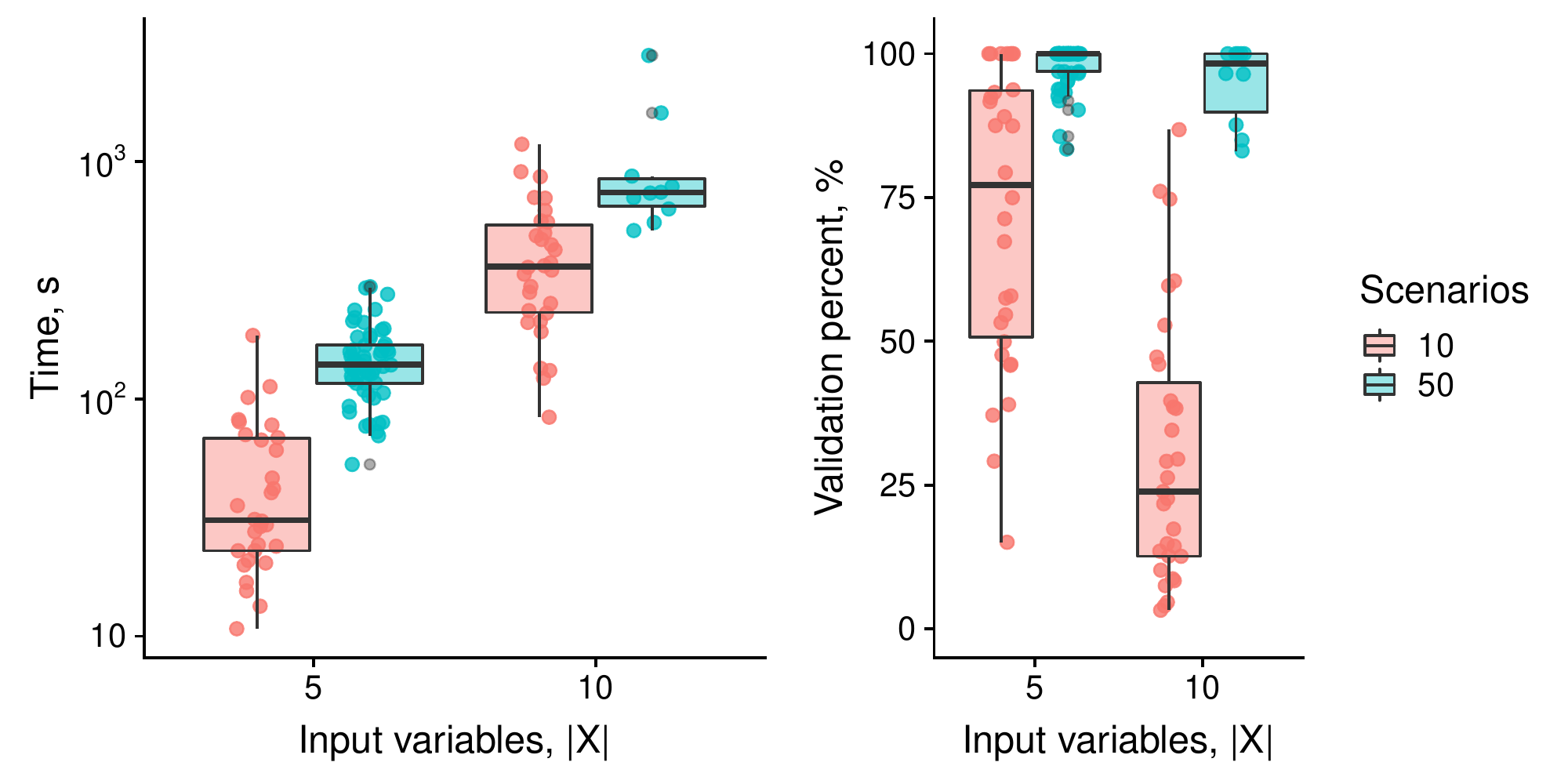}
    \vskip-\abovecaptionskip
    \caption{Distribution plots for Random Automata Case Study}%
    \label{fig:results-random}
    \vskip-\belowcaptionskip
\end{figure}

    \begin{figure}[p]
        \centering
        \includegraphics[width=\textwidth,totalheight=\textheight-2\baselineskip,keepaspectratio]{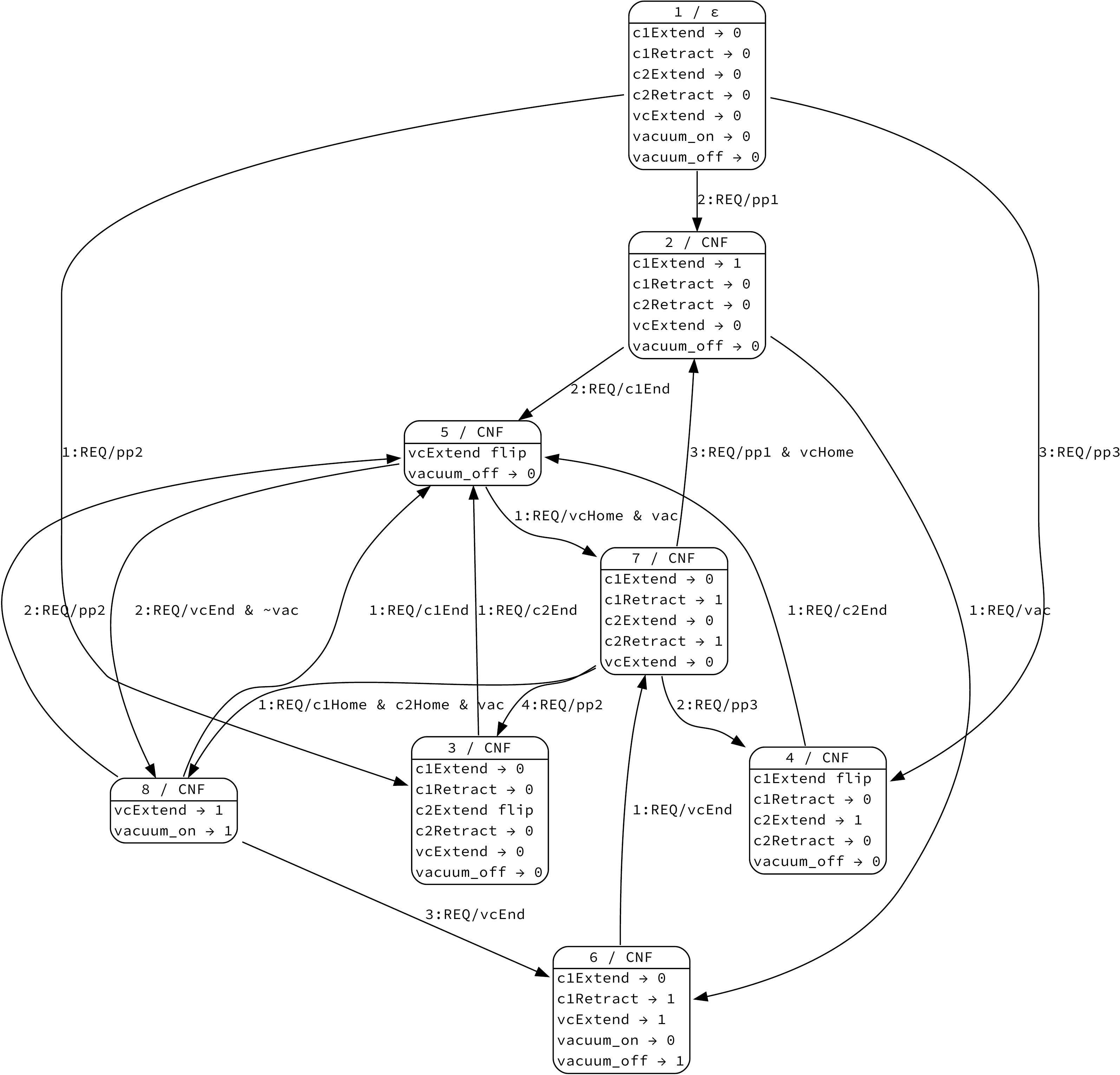}
        \caption{Example automaton generated using $\textsc{Complete*-min-CEGIS}$ algorithm from $\SetScenarios^{(39)}$; \enquote{flip} indicates a flip of corresponding output variable}
        \label{fig:example-automaton-complete-min-cegis-tests-39}
    \end{figure}

    \end{subappendices}
}{}


\begin{thebibliography}{10}
\providecommand{\url}[1]{\texttt{#1}}
\providecommand{\urlprefix}{URL }
\providecommand{\doi}[1]{https://doi.org/#1}

\bibitem{model-testing}
Apfelbaum, L., Doyle, J.: Model based testing. In: Softw. Qual. Week Conf. pp.
  296--300 (1997)

\bibitem{petrenko}
Avellaneda, F., Petrenko, A.: {FSM} inference from long traces. In: Formal
  Methods. pp. 93--109. Springer (2018)

\bibitem{sat-cardinality}
Bailleux, O., Boufkhad, Y.: Efficient {CNF} encoding of boolean cardinality
  constraints. In: Principles and Practice of Constraint Programming. pp.
  108--122. Springer Berlin Heidelberg, Berlin, Heidelberg (2003)

\bibitem{onehot-binary}
Bj{\"o}rk, M.: Successful sat encoding techniques. JSAT  \textbf{7},  189--201
  (2011)

\bibitem{buzhinsky-tii}
Buzhinsky, I., Vyatkin, V.: Automatic inference of finite-state plant models
  from traces and temporal properties. IEEE Trans. Ind. Informat.
  \textbf{13}(4),  1521--1530 (2017)

\bibitem{g4ltl-st}
Cheng, C.H., Huang, C.H., Ruess, H., Stattelmann, S.: {G4LTL-ST}: Automatic
  generation of {PLC} programs. In: Computer Aided Verification. pp. 541--549.
  Springer International Publishing, Cham (2014)

\bibitem{chivilikhin-18}
Chivilikhin, D., Buzhinsky, I., Ulyantsev, V., Stankevich, A., Shalyto, A.,
  Vyatkin, V.: Counterexample-guided inference of controller logic from
  execution traces and temporal formulas. In: 23rd IEEE International
  Conference on Emerging Technologies and Factory Automation. pp. 91--98 (2018)

\bibitem{fbCSP}
Chivilikhin, D., Ulyantsev, V., Shalyto, A., Vyatkin, V.: {CSP}-based inference
  of function block finite-state models from execution traces. In: 2017 {IEEE}
  15th International Conference on Industrial Informatics ({INDIN}). pp.
  714--719 (2017)

\bibitem{chivilikhin-19}
{Chivilikhin}, D., {Ulyantsev}, V., {Shalyto}, A., {Vyatkin}, V.: Function
  block finite-state model identification using {SAT} and {CSP} solvers. IEEE
  Transactions on Industrial Informatics  \textbf{15}(8),  4558--4568 (Aug
  2019). \doi{10.1109/TII.2019.2891614}

\bibitem{incremental-cryptominisat}
Chukharev, K.: Wrapper for incremental {SAT} solving using {Cryptominisat},
  \url{https://github.com/Lipen/incremental-cryptominisat}

\bibitem{fbSAT-tool}
Chukharev, K., Chivilikhin, D.: {fbSAT} tool,
  \url{http://www.github.com/ctlab/fbSAT}

\bibitem{fbSAT-full}
Chukharev, K., Chivilikhin, D.: {fbSAT}: Automatic inference of minimal
  finite-state models of function blocks (2019),
  \url{https://arxiv.org/abs/1907.03285}

\bibitem{NuSMV}
Cimatti, A., Clarke, E., Giunchiglia, F., Roveri, M.: {NuSMV}: a new symbolic
  model checker. International Journal on Software Tools for Technology
  Transfer  \textbf{2}(4),  410--425 (2000)

\bibitem{clarke1999model}
Clarke, E.M., Grumberg, O., Peled, D.: Model checking. MIT press (1999)

\bibitem{regular-inference}
Coste, F., Nicolas, J.: Regular {I}nference as a graph coloring problem. In: In
  Workshop on Grammar Inference, Automata Induction, and Language Acquisition
  (ICML' 97. pp.~9--7 (1997)

\bibitem{dubinin-2006}
Dubinin, V., Vyatkin, V.: Towards a formal semantic model of {IEC}~61499
  function blocks. In: IEEE Int. Conf. Ind. Informat. pp. 6--11 (2006)

\bibitem{not-bosy}
Faymonville, P., Finkbeiner, B., Rabe, M.N., Tentrup, L.: Encodings of bounded
  synthesis. In: Tools and Algorithms for the Construction and Analysis of
  Systems. pp. 354--370 (2017)

\bibitem{bosy}
Faymonville, P., Finkbeiner, B., Tentrup, L.: {BoSy}: An experimentation
  framework for bounded synthesis. In: Computer Aided Verification. pp.
  325--332. Springer, Cham (2017)

\bibitem{bounded-cycle}
Finkbeiner, B., Klein, F.: Bounded cycle synthesis. In: Computer Aided
  Verification. pp. 118--135. Springer International Publishing, Cham (2016)

\bibitem{giantamidis-tripakis}
Giantamidis, G., Tripakis, S.: Learning {Moore} machines from input-output
  traces. In: Formal Methods. pp. 291--309. Springer, Cham (2016)

\bibitem{gold}
Gold, M.: Complexity of automaton identification from given data. Information
  and Control  \textbf{37}(3),  302--320 (1978)

\bibitem{heule2010}
Heule, M., Verwer, S.: Exact {DFA} identification using {SAT} solvers. In: Int.
  Colloquium Conf. on Grammatical Inference. pp. 66--79 (2010)

\bibitem{klenze}
Klenze, T., Bayless, S., Hu, A.J.: Fast, flexible, and minimal {CTL} synthesis
  via {SMT}. In: Computer Aided Verification. pp. 136--156. Springer
  International Publishing, Cham (2016)

\bibitem{RINGA}
Lee, E., Kim, Y.G., Seo, Y.D., Seol, K., Baik, D.K.: {RINGA}: Design and
  verification of finite state machine for self-adaptive software at runtime.
  Information and Software Technology  \textbf{93},  200--222 (2018)

\bibitem{TESTOR}
Marsso, L., Mateescu, R., Serwe, W.: {TESTOR}: A modular tool for on-the-fly
  conformance test case generation. In: TACAS 2018. pp. 211--228. Springer,
  Cham (2018)

\bibitem{strix}
Meyer, P.J., Sickert, S., Luttenberger, M.: Strix: Explicit reactive synthesis
  strikes back! In: Computer Aided Verification. pp. 578--586. Springer
  International Publishing, Cham (2018)

\bibitem{neider}
Neider, D., Topcu, U.: An automaton learning approach to solving safety games
  over infinite graphs. In: Tools and Algorithms for the Construction and
  Analysis of Systems. pp. 204--221. Springer Berlin Heidelberg, Berlin,
  Heidelberg (2016)

\bibitem{nxtstudio}
{nxtControl} - {nxtStudio}, \url{http://www.nxtcontrol.com/en/engineering}

\bibitem{patil-pnp}
Patil, S., Vyatkin, V., Sorouri, M.: Formal verification of intelligent
  mechatronic systems with decentralized control logic. In: IEEE Conf. Emerg.
  Technol. Factory Autom. pp.~1--7 (2012)

\bibitem{petrenko2}
Petrenko, A., Avellaneda, F., Groz, R., Oriat, C.: Fsm inference and checking
  sequence construction are two sides of the same coin. Software Quality
  Journal  (2018)

\bibitem{rosner-phd}
Rosner, R.: Modular synthesis of reactive systems (1992), {PhD thesis.}

\bibitem{smetsers-lata}
Smetsers, R., Fiter{\u{a}}u-Bro{\c{s}}tean, P., Vaandrager, F.: Model learning
  as a satisfiability modulo theories problem. In: Klein, S.T.,
  Mart{\'i}n-Vide, C., Shapira, D. (eds.) Language and Automata Theory and
  Applications. pp. 182--194. Springer International Publishing, Cham (2018)

\bibitem{solar-lezama-2006}
Solar-Lezama, A., Tancau, L., Bodik, R., Seshia, S., Saraswat, V.:
  Combinatorial sketching for finite programs. SIGOPS Oper. Syst. Rev.
  \textbf{40}(5),  404--415 (2006). \doi{10.1145/1168917.1168907}

\bibitem{cryptominisat}
Soos, M., Nohl, K., Castelluccia, C.: Extending {SAT} solvers to cryptographic
  problems. In: Theory and Applications of Satisfiability Testing. pp. 244--257
  (2009)

\bibitem{tsarev-egorov-gecco}
Tsarev, F., Egorov, K.: Finite state machine induction using genetic algorithm
  based on testing and model checking. In: Conf. Comp. Genetic Evol. Comput.
  pp. 759--762. ACM (2011)

\bibitem{efsm-tools}
Ulyantsev, V., Buzhinsky, I., Shalyto, A.: Exact finite-state machine
  identification from scenarios and temporal properties. International Journal
  on Software Tools for Technology Transfer  \textbf{20}(1),  35--55 (2018)

\bibitem{ulyantsev-lata}
Ulyantsev, V., Zakirzyanov, I., Shalyto, A.: {BFS}-based symmetry breaking
  predicates for {DFA} identification. In: Language and Automata Theory and
  Applications. pp. 611--622. Springer, Cham (2015)

\bibitem{ulyantsev-efsm-sat}
Ulyantsev, V.I., Tsarev, F.N.: Extended finite-state machine induction using
  {SAT}-solver. IFAC Proceedings Volumes  \textbf{45}(6),  236 -- 241 (2012),
  14th IFAC Symposium on Information Control Problems in Manufacturing

\bibitem{closed-loop}
Vyatkin, V., Hanisch, H.M., Pang, C., Yang, C.H.: Closed-loop modeling in
  future automation system engineering and validation. IEEE Transactions
  on~Systems, Man, and Cybernetics, Part C: Applications and Reviews
  \textbf{39}(1),  17--28 (2009)

\bibitem{vyatkin-tii}
Vyatkin, V.: {IEC}~61499 as enabler of distributed and intelligent automation:
  State-of-the-art review. IEEE Trans. Ind. Informat.  \textbf{7}(4),  768--781
  (2011)

\bibitem{walkinshaw}
Walkinshaw, N., Taylor, R., Derrick, J.: Inferring extended finite state
  machine models from software executions. Empirical Software Engineering
  \textbf{21}(3),  811--853 (2016)

\bibitem{direct-encoding}
Walsh, T.: Sat v csp. In: Proceedings of the 6th International Conference on
  Principles and Practice of Constraint Programming. p. 441–456. CP ’02,
  Springer-Verlag, Berlin, Heidelberg (2000)

\bibitem{zakirzyanov2019}
Zakirzyanov, I., Morgado, A., Ignatiev, A., Ulyantsev, V., Marques-Silva, J.:
  Efficient symmetry breaking for {SAT}-based minimum {DFA} inference. In:
  Language and Automata Theory and Applications. pp. 159--173. Springer
  International Publishing, Cham (2019)

\end{thebibliography}
\end{document}